\documentclass[sn-vancouver,Numbered]{sn-jnl}

\usepackage{array}
\usepackage{braket} 
\usepackage{amsfonts} 
\usepackage{upgreek}
\usepackage{float}
\usepackage{graphicx}%
\usepackage{multirow}%
\usepackage{amsmath,amssymb,amsfonts}%
\usepackage{amsthm}%
\usepackage{mathrsfs}%
\usepackage[title]{appendix}%
\usepackage{textcomp}%
\usepackage{manyfoot}%
\usepackage{booktabs}%
\usepackage{algorithm}%
\usepackage{algorithmicx}%
\usepackage{algpseudocode}%
\usepackage{listings}%
\usepackage{setspace}%
\usepackage{hyperref}
\usepackage{caption}
\usepackage{booktabs}
\usepackage[table,xcdraw]{xcolor}
\usepackage{subcaption}
\usepackage{tabularx}
\usepackage{acronym}
\usepackage[utf8]{inputenc}  

\begin{document}

\singlespacing

\title[Article Title]{Exploring the mechanisms of qubit representations and introducing a new category system for visual representations: Results from expert ratings}


\author*[1,2,3]{\fnm{Linda} \sur{Qerimi}}\email{linda.qerimi@physik.uni-muenchen.de}
\author[4]{\fnm{Sarah} \sur{Malone}}
\author[5]{\fnm{Eva} \sur{Rexigel}}
\author[2,3]{\fnm{Sascha} \sur{Mehlhase}}
\author[1]{\fnm{Jochen} \sur{Kuhn}}
\author[1]{\fnm{Stefan} \sur{K\"uchemann}}

\affil[1]{\orgdiv{Chair of Physics Education, Faculty of Physics}, \orgname{Ludwig-Maximilians-Universität M\"unchen (LMU)}, \street{Geschwister-Scholl-Platz 1}, \postcode{80539} \city{M\"unchen}, \country{Germany}}

\affil[2]{\orgdiv{Max Planck Institute of Quantum Optics}, \postcode{85748} \city{Garching}, \country{Germany}}

\affil[3]{\orgdiv{Munich Quantum Valley (MQV)}, \postcode{80807} \city{M\"unchen}, \country{Germany}}

\affil[4]{\orgdiv{Saarland University, Department of Education}, \postcode{66123} \city{Saarbr\"ucken}, \country{Germany}}

\affil[5]{\orgdiv{Department of Physics, University of Kaiserslautern-Landau}, \postcode{67663} \city{Kaiserslautern}, \country{Germany}}


\abstract{In quantum physics education, it is essential to use representations such as diagrams, models, and visual aids that students can readily relate to mathematical concepts. Research in representation theory has shown that integrating symbolic-mathematical elements (e.g., formulae) with visual-graphical representations enhances conceptual understanding more effectively than representations that merely illustrate phenomena. Given that common representations vary widely in their characteristics and that existing categorisation systems fail to adequately differentiate between these representations in quantum physics, we developed a new set of differentiation criteria. These criteria are grounded in current insights from representation research, quantum education, and specific aspects of quantum sciences and technologies. To create a comprehensive category system for evaluating visual quantum physics representations for educational purposes, we adopted Ainsworth’s (2006) Design, Functions, and Tasks Framework as our conceptual foundation. 

Twenty-one experts in quantum physics from four countries evaluated the category system using four widely recognised qubit representations: the Bloch sphere, the Circle Notation, the Quantum Bead, and the pie chart (Qake) model. This evaluation allowed us to assess the discriminative power of our criteria and the effectiveness of each representation in supporting the learning of quantum physics concepts. The experts assessed the four types of representations across 16 criteria to determine how well each conveyed quantum concepts, including quantum state, quantum measurement, superposition and probability, entanglement, and quantum technologies (X-, Z-, and H-gates). 

The results revealed significant differences in the effectiveness of these representations, particularly in their ability to convey fundamental concepts like superposition, quantum measurement, and probabilistic aspects. Moreover, our analysis identified significant differences in expert ratings of the four representations regarding their potential to evoke misconceptions. These differences were primarily linked to variations in shape, measurement behaviour, and the fundamental requirements for understanding entanglement. In analysing these expert opinions, we discuss key considerations for the use and development of new representations in education and offer suggestions for generating hypotheses for future empirical studies.}


\keywords{quantum technologies, representations, qubit, quantum education, expert rating, category system}

\maketitle

\section{Introduction}\label{sec1}

\subsection{Quantum education}\label{subsec1.1}

In light of the growing importance of quantum technology (QT), the teaching of quantum physics (QP) plays a central role in various stages of education and professional training (school, university, and industry). It is vital to ensure that future generations are adequately prepared to meet the challenges and opportunities associated with the increasing use of these technologies.

Introducing the fundamentals of QT, in particular the qubit, not only leads to new approaches in the teaching of QP but also opens up the possibility of an application-oriented teaching methodology~\cite{passante_energy_2016, dur_was_2012, dur_was_2014}. In addition to the different approaches that can be used to introduce learners to QT, a recently updated competence framework addresses the understanding of QT at different levels and in different relevant sectors~\cite{greinert_future_2023, greinert_qualification_2022}. 

The spin-first approach appears to be a suitable method for teaching the fundamentals of QP and, with its focus on qubits as two-level systems, also for teaching QT~\cite{sadaghiani_spin_2016, dur_was_2012, dur_was_2014}. This approach introduces the spin-half context with the qubit at an early stage, providing a basic understanding of QP and, in particular, QT. The \textit{reasoning tools} for QP, also called \textit{the basic rules of quantum physics,} from the German "Wesenz\"uge der Quantenphysik"~\cite{kublbeck2002wesenszuge, muller_milqquantum_2021, MüllerGreinert+2024}, include the following topics that also can be addressed with the spin-first approach:

\begin{itemize}
    \item Quantum measurement
    \item Complementarity
    \item Indeterminism and statistical predictability
    \item Interference of single quantum objects
\end{itemize}

Current research has highlighted that these key concepts, in addition to superposition and entanglement, are identified by educators as fundamental to the teaching of QP and QT. In particular, superposition and entanglement are key to understanding QT~\cite{merzel_core_2024}. Sadaghiani et al. found that students demonstrated a higher understanding of QP concepts when they followed the spin-first approach compared with a first-position approach~\cite{sadaghiani_spin_2015}. 

To convey the complex content of QP in an intuitive and memorable way, it is essential to create an appropriate learning environment that encompasses not only an effective teaching approach but also the use of suitable representations. The selection of representations in the context of QP is a challenging endeavour. Stadermann elucidates the complexities in identifying useful representations in QP~\cite{stadermann_analysis_2019}: 

\textit{"In contrast to most classical physics topics, we cannot find a consistent visualisation for quantum phenomena. QP offers students new views on physical reality, which conflict with earlier learnt classical concepts such as the nature of particles, locality, and determinism. Scientists still discuss how -- and if at all -- QP should be interpreted."}.

Therefore, it is essential to identify valuable elements of visual representations within the context to generate hypotheses regarding their learning effectiveness and, potentially, to generate new representations. 

\subsection{Qubit representations}\label{subsec1.2}

The term “representation” can be interpreted in many different ways. There is a diversity of perspectives, definitions, and categorisations of representations in the context of education. In this work, we are orientated towards Lemke~\cite{lemke_multiplying_nodate}, who analysed the construction and conveyance of meanings through signs and symbols, both in verbal and non-verbal languages. He found that in scientific papers, content such as text, graphs, tables, photos, and equations are used together because meaning is constructed through a multimodal process. This meaning is now taken up for learning. To determine the conditions in which this multimodality can be employed in the context of learning QP and QT, we categorise possible representations. Examples of different types of representations that could be used and combined in multimodal quantum instruction are shown in Table \ref{tab_mulitmodal}.

\begin{table}[!ht]
\centering
\caption{Example for representation in different variations}
\label{tab_mulitmodal}
\begin{tabular}{|p{6cm}|p{6cm}|}
\hline
\textbf{Description} & \textbf{Example} \\
\hline
Visual-graphical 

(e.g. diagrams, graphs, images) & \begin{center}\includegraphics [width=2cm]{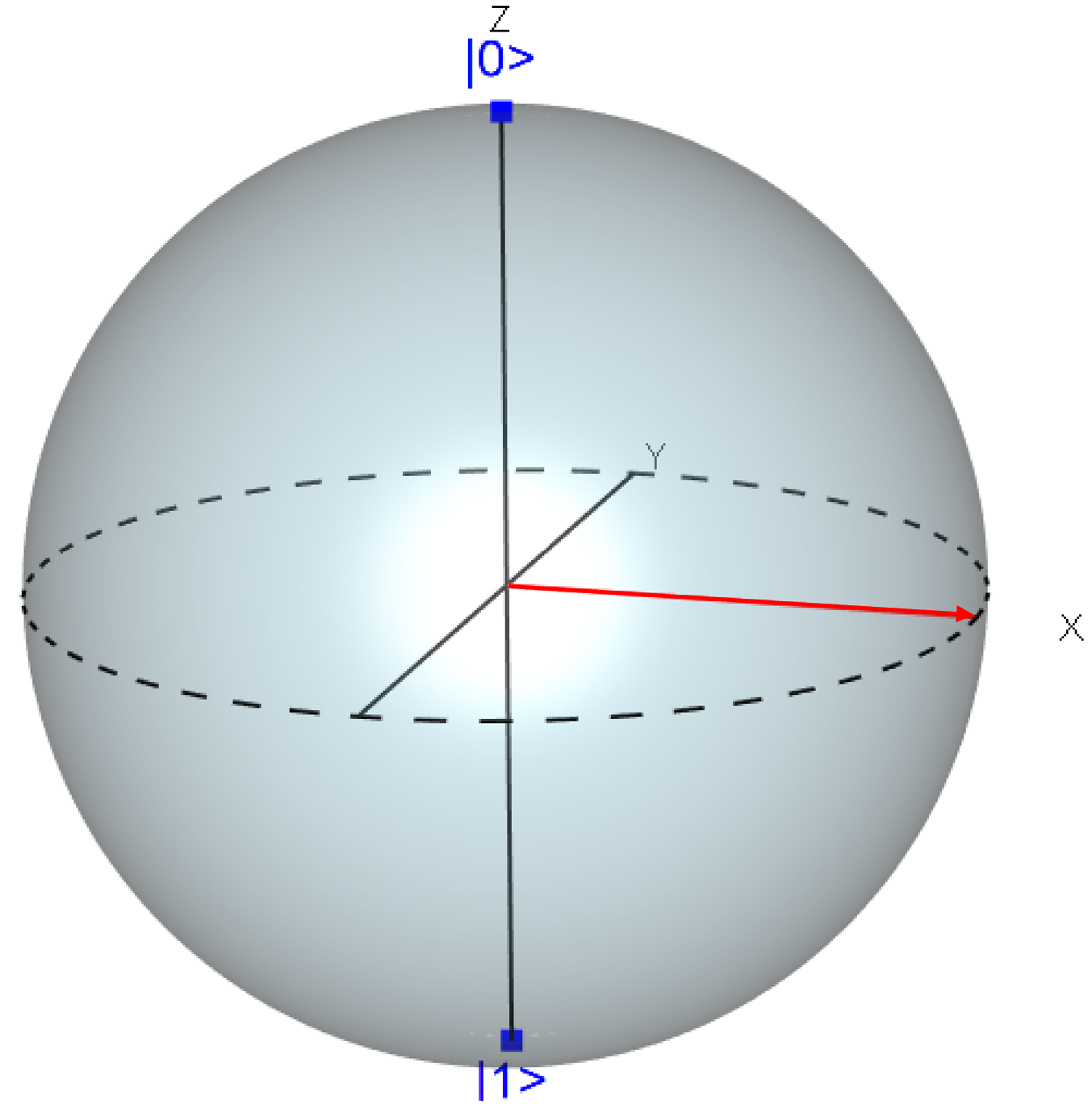} 

Bloch sphere \end{center}\\
\hline
Mathematical (-operational) 

(e.g. formulae) & \begin{equation}
\ket{\Psi} = a_1 \ket{0} + a_2 \ket{1}
\end{equation}

{\small {Remark:} $a_{1,2} \in \mathbb{C}$, \quad $\sum_{i=1}^{2} |a_i|^2 = 1$.} 

\\
\hline
Visual-gestural 

(e.g. physical or gestural movements)  & \begin{center}\includegraphics [width=1.5cm]{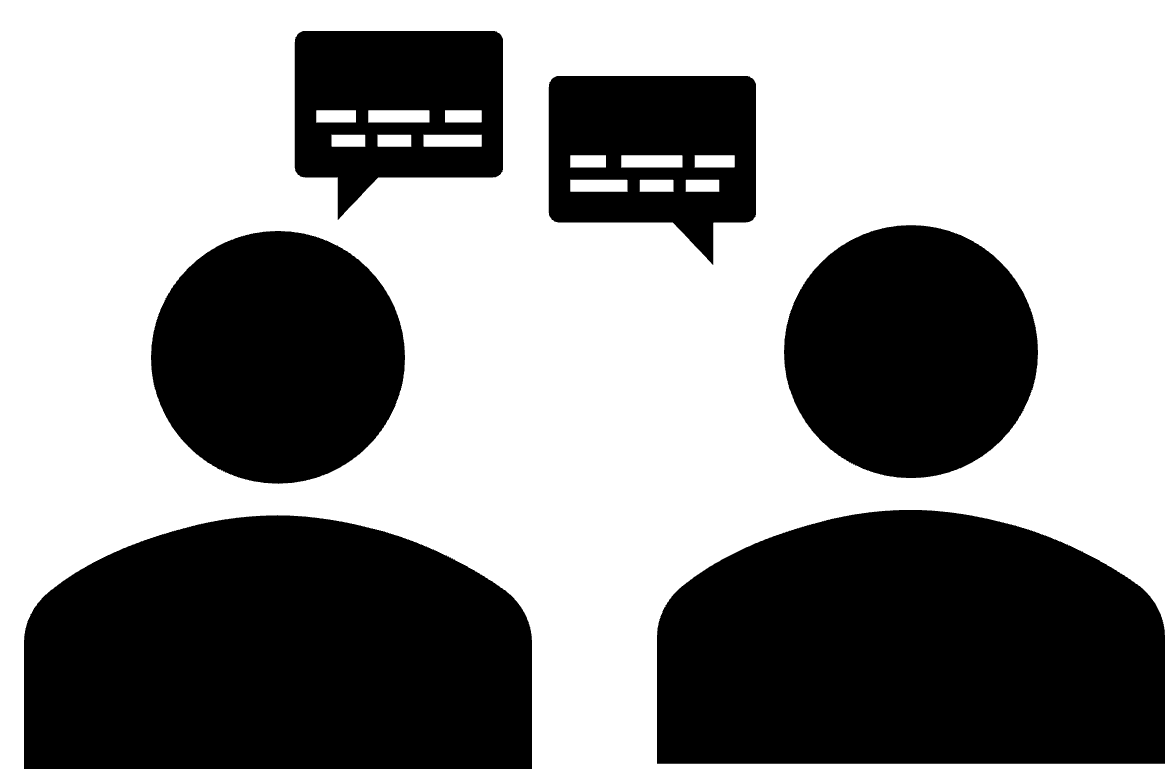} \end{center}

Interaction between the teacher and the student through physical and gestural movement 

\\
\hline
Verbal-semantic 

(e.g. Text) & 
"Superposition can be realised with the aid of a beam splitter." 
\\
\hline
\end{tabular}

\end{table}

Previous research has already attempted to implement various representations in combination with the spin-first approach~\cite{manogue_representations_2012}. One commonly used visual-graphical representation of two-level systems is the Bloch sphere, which has several variants ~\cite{woitzik_quanteninformationsverarbeitung_2020,  dur_was_2012, dur_was_2014}. However, a variety of other representations are also available, for example, the “arrow” formalism by Richard Feynman~\cite{feynman2006qed, bader1994optik, kublbeck1997modellbildung, kublbeck_quantenphysik_2015} and the Circle Notation~\cite{johnston_programming_2019, bley_visualizing_2024, just_quantencomputing_2020}. In recent years, many other visualisations of qubits have been developed and refined, for example, the Quantum Bead by Steffen Glaser's group~\cite{garon_visualizing_2015} (second paper in progress), and the pie chart model~\cite{yeung2020quantum}: Qubit cake model (Qake, paper in progress)~\cite{Donhauser2024}. For the visualisation of quantum phenomena, not only have different representations been used and analysed, but also media such as the quantum Composer~\cite{weidner2021publications, kuchemann_impact_2023} or the Quantum Mechanics Visualisation Project (QuVis)~\cite{kohnle_sketching_2020, kohnle_investigating_2015}. The field of visual-graphical quantum representation is undergoing constant development and refinement. There is a significant interest in making QP accessible to an interdisciplinary audience, particularly with regard to the future development of QT~\cite{goorney_quantum_2023}.    

To effectively teach QP and QT to learners with little or no prior knowledge, it is crucial to impart the necessary mathematical skills. For a deeper understanding of QP and QT concepts, it is recommended to become familiar with the relevant mathematical formalism~\cite{greinert_qualification_2022}. Given that the formalism of QP can be challenging to comprehend and apply, visual-graphical representations of qubits can be used to facilitate understanding. Learning often involves the use of multiple resources, such as text, interactions, mathematical formulae, and visual-graphical representations, such as the Bloch sphere. It is crucial that educators carefully evaluate and analyse qubit representations and their properties to ensure their suitability for conveying the key concepts. This involves verifying that the visual representations clearly reflect the underlying mathematical principles. Concurrently, learners need to engage with these diverse representations proactively by drawing connections between them. This interaction helps students grasp the mathematical concepts and reinforces the structural relationships within QP. Connecting multiple visual representations requires identifying relevant similarities and understanding the conventions that guide their integrated use. This is referred to as \textit{connectional understanding}~\cite{rau_conditions_2017}. 

The exclusive focus on qubit representations within the spin-first approach provides the opportunity to visualise and elucidate the fundamental principles of QP with suitable representations. 

Although it has been suggested that visual-graphical qubit representations are beneficial due to their accessibility and close relationship with mathematics, their perceived simplicity can pose a risk that they will be misunderstood. Learners sometimes fail to recognise that they are not direct representations of reality, but scientific models. A prominent class of such difficulties is the “graph-as-picture” misconception, in which, for example, a learner may misinterpret a line graph as a picture of a mountain~\cite{garcia_garcia_graph-as-picture_2010}. Therefore, their risk of supporting the development of misconceptions (e.g. the Bloch sphere describes the behaviour of a photon, then the photon is associated with the shape of the Bloch sphere as a small sphere) should be considered when using or creating visual-graphical representations.

With regard to the use of qubit representations, difficulties and misconceptions in QP have been analysed extensively by various authors~\cite{wiesner1996verstandnisse, fischler_modern_1992, singh_review_2015}. The following are selected misconceptions identified in the literature regarding the concepts of quantum state, quantum measurement, superposition, and entanglement that may be encountered when using visual-graphical representations:

\begin{enumerate}
    \item Difficulties with measurement and expected value have been identified~\cite{singh_review_2015}. It should be noted that the representation is not understood as an object that predetermines the measurement but as a way of predicting measurement results (under certain probabilities).
    \item In a study on mental models, it was found that the majority of participants held the misconception of "spin as a rotation of particles around their own axis"~\cite[p. 1374]{ozcan_what_2011}. When selecting and using qubit representations, this should be taken into account to avoid activating or reinforcing misconceptions associated with the visual properties (e.g. shape) or function (e.g. rotation) of the representations. 
    \item It is difficult for students to understand that photons can be in two states (e.g. horizontal or vertical polarisation). The majority of students encountered difficulty in accepting that the polarisation states of a photon can be employed as the basis for a two-state system~\cite{singh_review_2015}. This was frequently observed in students who exhibited a pronounced inclination towards their established understanding of polarisation within the context of classical optics~\cite{singh_review_2015}. The discrepancy between the visualisation of polarisation states in a two-state system may be reinforced by the use of visual-graphical representations of, for example, the behaviour of a photon after a beam splitter.
    \item Using qubit representations, it should be possible to connect to more complex topics such as entanglement and complex quantum gates. To our knowledge, no authors have identified specific misconceptions caused by representations of entanglement; nevertheless, it is important to ensure that they convey the abstract concept without causing further difficulties. We have limited ourselves to the fundamental requirements for understanding entanglement: going from a one-qubit to a two-qubit system. The focus was on clarifying the visual-graphical representations, which show the necessity of at least two qubits for entanglement. However, the limitations of the representations should not be seen as reasons to exclude them. They can certainly be used for educational purposes, but a transition with multiple representations should be able to fill this “gap” and provide a comprehensive representation of quantum concepts, especially those that are essential for QT.
\end{enumerate}

When creating or selecting visual-graphical representations, care should be taken to ensure that they are designed in a way that best supports concept acquisition and does not create misconceptions. However, the manner in which a representation is presented can also lead to difficulties~\cite{henriksen_what_2018, ubben_two_2022}.

\subsection{Describing visual-graphical qubit representations}\label{subsec1.3}

As in other areas of physics, educators in QP are confronted with the challenge of comparing and selecting appropriate representations. To assist them, it is essential to categorise representations and describe them in a consistent manner to evaluate their strengths, weaknesses, and special features. Previous categorisations of representations, such as those proposed by Lemke~\cite{lemke_multiplying_nodate}, Schnotz~\cite{schnotz_kognitive_2001}, Kosslyn~\cite{kosslyn_understanding_1989} and Bertin~\cite{bertin1983semiology}, differentiate visual-graphical representations but do not support drawing conclusions about their effectiveness or appropriate use. However, if educators could gain a deeper understanding of the aspects of qubit representations that promote the acquisition of content knowledge, they could assign qubit representations to different levels of learners’ prior knowledge and develop more effective, targeted approaches.

In this work, we present a refined categorisation system for representation and QP visualisation research. We selected Ainsworth’s Design, Functions, and Tasks (DeFT)~\cite{ainsworth_deft_2006} as the conceptual framework and extended it with relevant aspects of QP representations, including their respective potential risks of inducing misconceptions in learners. Independent of the learning content, the DeFT Framework provides an overview of how multiple external representations (MER) can be used effectively to support students' learning~\cite{ainsworth_deft_2006}. Ainsworth outlined relevant aspects of design, functions, and tasks when learning with MER. The design aspect of the DeFT framework encompasses the features of multirepresentational learning environments that directly influence learner interaction; it relates to accessibility, comprehensibility, active engagement, and cognitive connections across different representations. The “functions” refer to the roles that can be played by multiple representations in supporting learning: providing complementarity, constraining interpretation, and constructing deeper understanding. These functions influence how learners process and integrate information. The “tasks” refer to the cognitive demands that learners must manage to work effectively with these representations; we have focused on content-related processes and operations that can be developed with the representations to solve tasks. Together, these dimensions provide an understanding of how incorporating MER into educational settings can influence learning processes and outcomes.

We decided to use DeFT~\cite{ainsworth_deft_2006} as a theoretical framework to derive useful dimensions for categorising visual representations (see \autoref{tab:categories_classification}).

Under \textbf{design}, categories were included that allow statements about the shape and visual impression of a representation. 

Under \textbf{function}, categories were included that primarily characterise the representations according to their interaction with the learners or other representations.

Under \textbf{task}, categories have been chosen that primarily take place in a basic QT application or task.

Finally \textbf{cross-concepts}, include aspects that do not fit in the categories above (the categories are described in more detail in Section  \ref{subsec2.3}).

\begin{table}[!ht]
\centering
\caption{Refined categorisation of visual representations}
\label{tab:categories_classification}
\begin{tabular}{>{\raggedright\arraybackslash} p{7cm} p{5cm}}
\toprule
\large\textbf{Category} & \large\textbf{Classification} \\
\midrule
\rowcolor{orange!30} \large 1. Salience & \\
\rowcolor{orange!30} \large 2. Dimension & \large Design\\
\rowcolor{orange!30} \large 3. Understanding difficulties & \\
\rowcolor{orange!30} \large 4. Colour & \\
\rowcolor{blue!20} \large 5. Actions/Steps & \\
\rowcolor{blue!20} \large 6. Interaction with mathematics & \\
\rowcolor{blue!20} \large 7. Contiguity & \large Functions \\
\rowcolor{blue!20} \large 8. Overlap/redundancy & \\
\rowcolor{blue!20}  \large 9. Complementarity & \\
\rowcolor{blue!20} \large 10. Predictability & \\
\rowcolor{green!20} \large 11. Phase visualisation & \\
\rowcolor{green!20} \large 12. Amplitude visualisation & \large Tasks/applications  \\
\rowcolor{green!20} \large 13. Concepts & \\
\rowcolor{green!20} \large 14. Quantum technology & \\
\rowcolor{gray!20} \large 15. Generability & \\
\rowcolor{gray!20} \large 16. Effort in explanation & \large Cross-concepts\\
\bottomrule
\end{tabular}
\end{table}
\subsection{Research questions}\label{subsec1.4}

Our goal is to refine a categorisation of representations that allows us to make decisions about the selection and design of visual representations for appropriate, effective, and sustainable learning of QP and QT content. We use expert ratings to profile and cluster representations and obtain answers to the research questions:

\textbf{RQ1:} Which features/aspects of visual-graphical qubit representations are relevant to learning quantum physics and quantum technology?

\textbf{RQ2:} What factors should be considered when creating new qubit representations to promote learning?

\section{Methods}\label{sec2}

\begin{figure}[ht!]
\centering
    \includegraphics[width=0.5\textwidth]{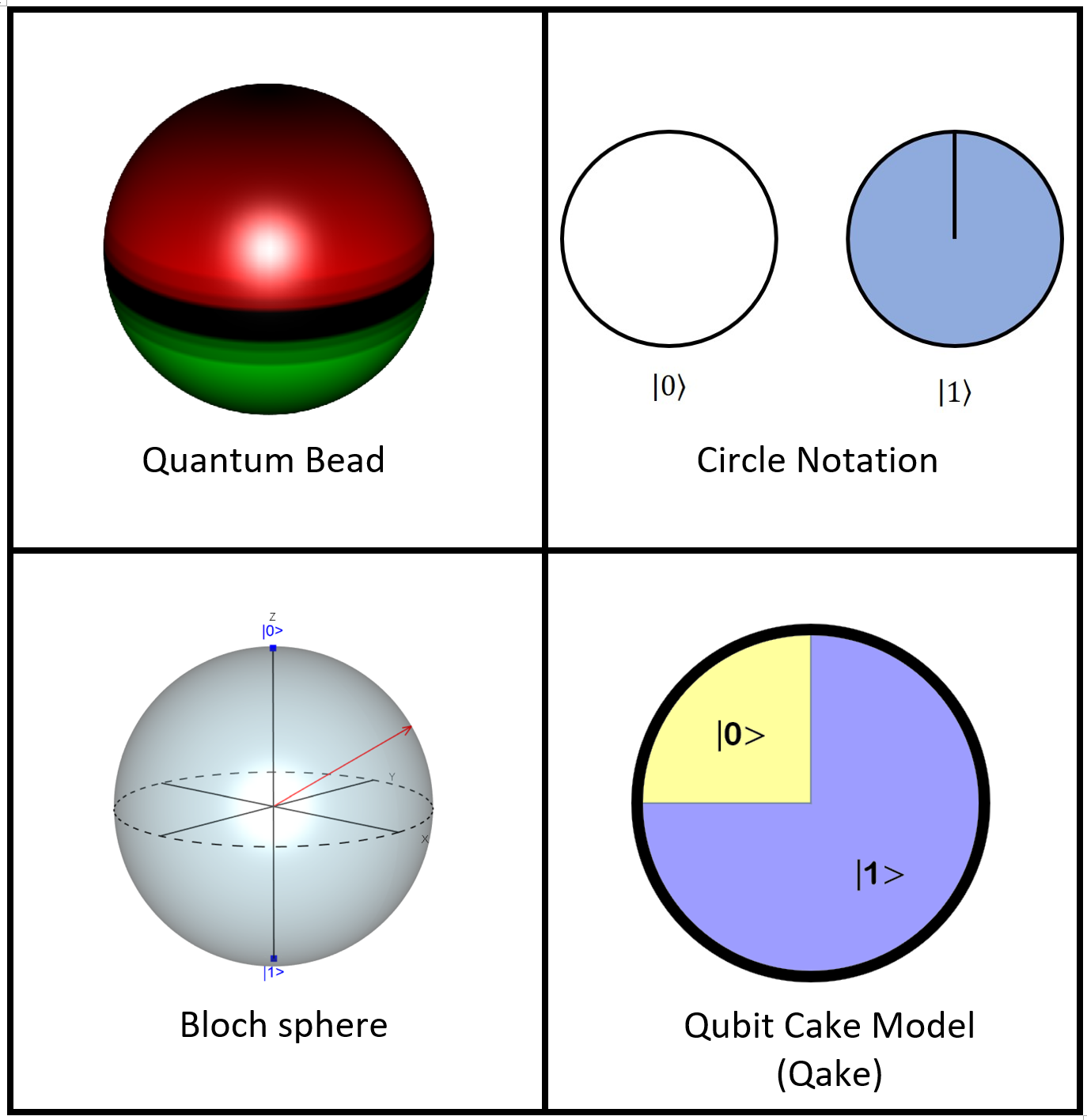}
    \caption{The Quantum Bead and Bloch sphere are three-dimensional representations while, whereas the Qubit Cake Model (Qake) and Circle Notation are two-dimensional representations. The Circle Notation uses the third dimension for multi-qubit systems. Own illustration based on Quantum Bead~\cite{leiner_wigner_2017, Huber2024}, Circle Notation~\cite{johnston_programming_2019}, the Bloch sphere~\cite{dur_visualization_2014} and Qake Model~\cite{yeung2020quantum, Donhauser2024}.}
    \label{img_fourQubit}
\end{figure}

To answer the research questions, we conducted online sessions in which experts were asked to rate four visual-graphical qubit representations across 16 categories using cheat sheets developed for the study. The Bloch sphere, the Cirle Noation~\cite{johnston_programming_2019, bley_visualizing_2024}, the Quantum Bead, a further development of the Spindrops representation~\cite{garon_visualizing_2015, Huber2024} (second paper in progress) and the pie-chart model: Qubit Cake Model (Qake)~\cite{Donhauser2024} were the four representations evaluated by the experts (\autoref{img_fourQubit}). The Quantum Bead is not substantially different from the Spin Drops representation (or DROPS for short~\cite{garon_visualizing_2015, leiner_wigner_2017}) for single qubits. The difference lies in the fact that the Quantum Bead is also suitable for visualising two or more qubits, including entanglement~\cite{Huber2024}, as requested in the expert rating. Therefore, we refer here to the Quantum Bead~\cite{Huber2024}.

\vspace{-2.0\baselineskip}
\subsection{Cheat sheets}\label{subsec2.1}
The structure and layout of the cheat sheets with the relevant terminology were identical for each of the four representations. Brief explanations were given on qubits (in general), quantum states (two-state system), the representation itself, and how the respective representations visualise the following: quantum measurement, superposition, entanglement and quantum gates (X-Gate, Z-Gate and H-Gate). The experts were advised first to look at the cheat sheets to familiarise themselves with all the qubit representations and then complete the rating sheet. The cheat sheets could optionally remain open. Experts were asked to assess each of the four representations within the different categories via Google Forms (\href{https://docs.google.com/forms}{https://docs.google.com/forms}). 

\subsection{Sample}\label{subsec2.2}
For the present study, we selected and invited experts whose current field of research is related to QT (theoretical, experimental, educational). Twenty-one experts from ten locations across four countries (Germany, Italy, Switzerland, and the USA) were involved. Only QT experts with teaching experience in QP were included (two professors or junior professors, nine postdoctoral researchers, and ten PhD students). PhD students were eligible to participate only if they were in their second year or above. The mean number of years spent engaged in research in QP was 5.1 years (standard deviation ±1.9 years). One participant with over 30 years of experience was considered an outlier and so was not included in this average. However, the ratings of this expert was considered for the analysis in this paper in the same way as the other data. The participants demonstrated expertise in various areas, including mathematics, quantum optics, quantum field theory, quantum computing, and quantum education. All experts were contacted personally or by email.  

\subsection{Categories}\label{subsec2.3}
The categories for evaluation were chosen based on Ainsworth's DeFT framework and research on (mis-)conceptions in quantum education and quantum sciences.
\begin{enumerate}
\item \textit{Salience}: This describes how clearly a concept is perceived through a representation. The salience of a stimulus can depend on its intensity, novelty, ecological validity, movement, and interactivity~\cite{hewstone2021introduction}. The signalling principle states that learning materials are more effective when they contain cues or elements that draw learners' attention to the relevant content or information or highlight the organisation of the content~\cite{mayer_signaling_2014}. 

\item \textit{Dimension}: Ainsworth mentions dimensionality as a relevant design factor~\cite{ainsworth_deft_2006}. Although visual-graphical representations are always two-dimensional, they can differ in whether and how three-dimensional information is visualised~\cite{ainsworth_deft_2006}. The capacity for spatial abilities, including mental rotation, differs between learners of different genders~\cite{castro-alonso_sex_2019, heo_learning_2020, saha_he_2016}.The visuospatial experience provided by representations depends on whether learners can adequately perform any necessary explicit operations (e.g. mental rotation). A study showed that spatial ability is the decisive factor for learning success, with male participants performing better than female participants regardless of the type of multimedia resources~\cite{heo_learning_2020}. Three-dimensional representations may require a higher level of spatial imagination than two-dimensional representations, that not all learners can achieve in the same way~\cite{heo_learning_2020}.  

\item \textit{Understanding difficulties}: This assesses whether the representations could lead to students misunderstanding the underlying concepts. Various topics are covered, including simple structures, quantum measurement, superposition, and entanglement. A more detailed background on misunderstandings that can arise from visual-graphical qubit representations is provided in Section \ref{subsec1.2}.

\item\textit{Colour:} Another component of representations is the \textit{colour} coding of concepts/information. The point here is not that colour can be used to visually draw attention, but that information is encoded with the colour, which can increase the extrinsic cognitive load and have a detrimental effect on learning ~\cite{sweller_cognitive_1994}. The additional information processing through colour requires increased (extrinsic) cognitive load and has a negative effect on learning. For example, both the colour and the mixture of colours (in the case of a function of the representation) contain information about the underlying content.

\item \textit{Actions/steps}: This category describes the cognitive steps required to extract relevant information from the representation of a concept or to perform a specific operation with the representation. It is based on Sweller’s element interactivity~\cite{sweller_cognitive_1994}. Learning content or representations that require numerous cognitive steps place a significant load on working memory, which is inherently limited in its capacity to process information effectively~\cite{baddeley_working_1992}. According to Miller, the human working memory can typically manage only a limited number of elements, often cited as $7 \pm 2$ at any given time~\cite{miller_magical_1956}. Therefore, representations should be as comprehensive as possible while remaining simple to avoid cognitive overload and instead provide support to learners. It is important to distinguish between extrinsic cognitive load, which arises from the way information is presented, and intrinsic cognitive load, which depends on the inherent complexity of the material relative to the learner's prior knowledge. Certain representations might require many cognitive steps for learners with low prior knowledge, leading to overload, yet be entirely suitable for learners with higher levels of expertise~\cite{sweller_cognitive_2019}. 

\item\textit{Mathematics:} For QP, the interaction between mathematics and visual-graphical representation is an important aspect in the choice of representation. The integration between mathematical and visual-graphical representations promotes conceptual understanding more than one representation that visualises certain phenomena~\cite{ainsworth_deft_2006, rau_conditions_2017}. Erwin Schr\"odinger captured the mathematical meaning of QP succinctly when he said \textit{“... then the mathematical apparatus of the new theory can give us a well-defined probability distribution for every variable ...”}~\cite{schrodinger_gegenwartige_1935} emphasising how fundamental a connectable representation is. To introduce mathematics at the right level or to enable later levels of learning, it is important to provide access to mathematics with visual-graphical representations.

\item \textit{Contiguity}: This is based on the principle of contiguity described by van Gog~\cite{mayer_signaling_2014}, who found that learners achieve greater learning success when text is placed next to the graphic~\cite{Mayer_Fiorella_2014}. The combination of both proved to be conducive to learning, which is why this category is included to consider the direct integration of a visual-graphical qubit representation with a mathematical formula or some additional text for a more detailed description. This could also ease the transition to the introduction of mathematical representations.  

\item \textit{Overlap/redudancy:} There is a paucity of data concerning the impact of redundant combinations of representations on learning outcomes. The combination of images and written text supports the acquisition of knowledge more effectively than the simultaneous presentation of the same information in the form of images, written text, and speech~\cite{Mayer_Fiorella_2014}. However, a study has demonstrated that combining multiple representations, especially text with formulae, improves problem-solving performance in mathematics, even if they contain redundant information~\cite{ott_multiple_2018}. Furthermore, the use of multiple visual-graphic representations can be beneficial for learning~\cite{ainsworth_deft_2006}. This includes the use of redundant graphical representations. A categorisation of representations in connection with one or more redundant representations enhances the use of MER and increases the potential to improve learning~\cite{ainsworth_deft_2006}. The challenge here is to identify compatible representations and use them in a targeted manner.   

\item \textit{Complementarity}: This describes the use of two visual-graphical representations that complement each other. The principle of complementarity implies that the representations differ either in the information they present or the related cognitive processes~\cite{ainsworth_deft_2006}. For example, two visual-graphical representations may depict the information of a superposition state while triggering different cognitive processes, thus promoting learning through stronger integration~\cite{rau_conditions_2017}. Another possibility is that one representation may depict the superposition well but not the entanglement, whereas the other representation may visualise the entanglement but not the superposition. Multiple representations can, therefore, be used to promote learning if they complement each other in their representation of concepts. As in \textit{overlap/redundancy}, the categorisation of representations in relation to one or more other complementary representations also has the potential to enhance the effectiveness of MER and improve learning~\cite{ainsworth_deft_2006}.

\item \textit{Predictability}: For representations to be used effectively by learners, they should enable predictions to be made. Thus, statements should be made about possible measurement results, taking into account the properties or rules of the representation~\cite{kruger_modelle_2018, treagust_multiple_2017}. Particularly important here are the tasks/applications categories, which capture how certain representations can be used to demonstrate specific applications or tasks of QT. These categories are specific to QT and the use of the qubit in the spin-first approach.

\item \textit{Phase (or phase change)}: This is a particularly important property of a quantum state. Global phases and relative phases were considered separately. The global phase refers to a phase-related transformation applied to all states of a quantum system. It has no direct influence on the observable phenomena of the system. Relative phase refers to the phase relationship between different states within a quantum system~\cite{nielsen_quantum_2010}.

\item\textit{Amplitude}: To analyse a two-state system in detail, terms such as “amplitude” and “phase” are needed ~\cite{feynman_quantum_2010, nielsen_quantum_2010}. To interpret and understand these terms as clearly as possible, it is important to visualise them.

\item \textit{Concepts:} The concepts (A) quantum measurement, (B) superposition, (C) entanglement, (D) probabilistics were considered to assess which concepts are adequately conveyed by the representations. The concepts were selected based on relevance of the spin-first process and the reasoning tools for QP, as well as the competence framework~\cite{sadaghiani_spin_2015, kublbeck2002wesenszuge, greinert_future_2023, MüllerGreinert+2024}.     

\item \textit{Quantum technologies}: This category focuses on direct application and captures which quantum gates are adequately conveyed by the representation of experimental components. This concerns common gates: the H-gate, the X-gate, and the Z-gate~\cite{garon_visualizing_2015}. 

\item \textit{Generability}: This refers to how readily a representation can be reproduced by learners. The difficulty or complexity required to create the representation should be assessed. For example, in a school context, we would consider how difficult/complex it is for learners to draw a representation in their exercise books when they are shown it on a blackboard~\cite{ainsworth_deft_2006}.

\item \textit{Effort in explanation}: This aims to categorise the effort required to explain a representation. For example, there are some representations that can illustrate many concepts, but the effort or complexity of explaining all these concepts may be so great that these representations are ineffective. 
\end{enumerate}

\subsection{Rating}\label{subsec2.4}

\begin{figure}[H]
    \includegraphics[width=1.0\textwidth]{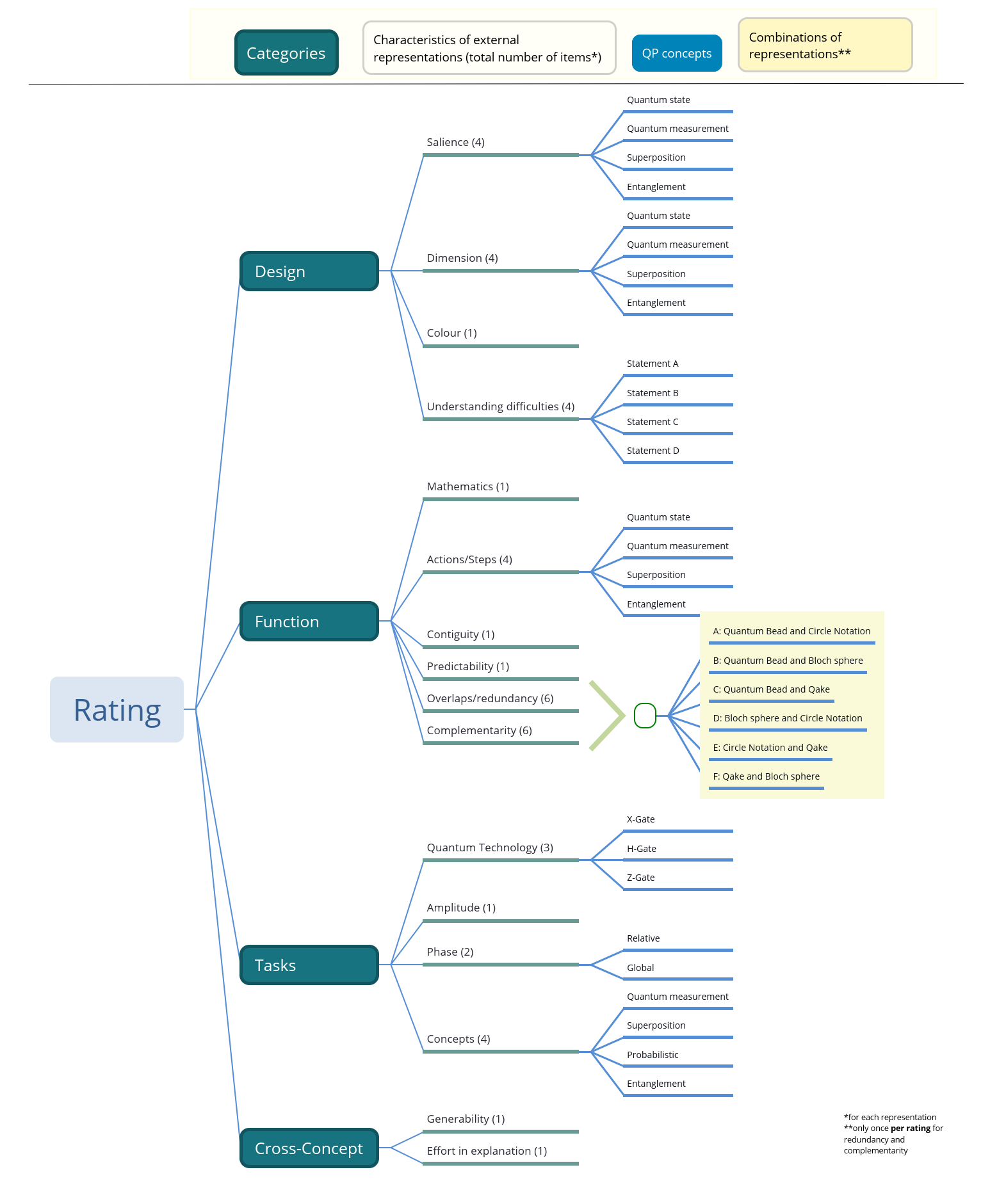}
    \caption{Diagram of the expert ratings: to evaluate the \textit{salience, dimension and actions/steps and concepts}, four items (for each QP concept) per representation are included. For \textit{understanding difficulties}, we include the items in statements A, B, C, and D. For \textit{redundancy and complementarity}, the combinations A–F are taken into account only once. The total number of items is calculated by adding the number of items on the sub-path of the characteristic of external representation, here QP concepts, so \textit{redundancy and complementary} are not taken into account as they are cross-representation items (32 items for one representation and 128 items for four). The number of all cross-representation items (Redundancy and Complementary) is shown here by 12 (6 + 6). The total number of items is calculated by multiplying the 32 items by 4 (for the representations) and adding 12 cross-representation items (into account once).}
    \label{Ablauf_Rating.png}
\end{figure}

The structure of the ratings shown in \autoref{Ablauf_Rating.png} indicates the weightings assigned to the different categories.

On a Likert scale from 1 to 5 (“strongly disagree” to “strongly agree” or the reverse), the experts were asked to assess the four qubit representations in each of the 16 categories. In addition to these options, the raters could select “I can’t judge, I do not know” for each item. Salience, dimension, and actions/steps were rated separately for the following concepts: \textit{Salience, Dimension and Action/Steps}, were rated separately for the concepts of 
\begin{itemize}
    \item quantum state
    \item quantum measurement
    \item superposition and probabilistic
    \item entanglement
\end{itemize}

We also measured how many actions/steps were required to extract the relevant information about the relevant concepts from each representation or to perform a specific operation with the representation. Next, we transformed this into a rating scale from 1 to 5 in increments of 0.5. The rating 5 was used for 1–2 steps and 4.5 for 3–4 steps, with this pattern continuing down to a rating of 2 for 13–14 steps. A rating of 1 was used if 15 or more steps were required. A lower number of steps means a lower number of cognitive processes, based on Sweller’s~\cite{sweller_cognitive_1988, sweller_cognitive_1994} concept of “element interactivity”. This term is used to describe the intrinsic cognitive load associated with the processing of different pieces of information or elements/aspects~\cite{sweller_cognitive_1994}. It encompasses the manner in which these elements interact with each other and how they are processed by the brain in order to be understood. A higher level of element interactivity is indicative of a greater (intrinsic) cognitive load, as it necessitates establishing more connections between the elements ~\cite{sweller_cognitive_1994}. 

In \textit{understanding difficulties}, the experts were asked to rate four possible statements (A–D) that might be made from the perspective of a students who may be unfamiliar with the respective representations. Based on works by a range of authors about mental models and misconceptions in QP~\cite{singh_review_2015, wiesner1996verstandnisse, fischler_modern_1992, marshman_investigating_2017, ozcan_what_2011, ubben_gestalt_2021}, the following statements were chosen:

\begin{description}
    \item A: "Based on the representation, I imagine the quantum object as a small sphere."
    \item B: "After a beamsplitter, the qubit will be split in two directions". 
    \item C: "The information of the qubit was already known \textit{before} the actual measurement and was confirmed with the representation (deterministic behaviour)." 
    \item D: "An entangled state is only possible with two or more qubits."
\end{description}
Statement D is correct and was included to capture difficulties in the transition to multi-qubit systems, with regard to the concept of entanglement. 

For \textit{complementarity} and \textit{overlaps/redundancy}, groups A–F were formed to evaluate the representations in relation to each other.  
\begin{description}
\item A: Quantum Bead and Circle Notation
\item B: Quantum Bead and Bloch sphere
\item C: Quantum Bead and Qake
\item D: Bloch sphere and Circle Notation
\item E: Circle Notation and Qake
\item F: Qake and Bloch sphere
\end{description}

The participants were asked to provide a separate rating for the visualisation of global and relative phases. For \textit{quantum technologies}, the rating was divided into the following topics: X-gate, Z-gate, and Hadamard gate (H-Gate). The experts gave their ratings for each gate. An attempt was made to limit the rating to simple variants of operations that could be experimentally transferred to optical elements~\cite{scholz_deutsch-jozsa_2006, obrien_demonstration_2003}. 

The 16 categories are taken into account, with four items to refine them in four concepts for \textit{salience, dimension, actions/steps} and also four items in \textit{concepts} to specify them, which (in general) appropriately convey the concepts. Four items were also included for \textit{understanding difficulties} because of the four "statements" describing interpretations or misinterpretations of concepts in QP. With \textit{phase} (2 items) and \textit{quantum technologies} (3 items), there are 32 items. Participants were asked about these for all four representations, which brings us to 128 items. Finally, the items were used in combinations due to \textit{overlaps/redundancy and complementarity}; here there are 12 items, bringing us in total to 140. 

In addition, the experts were asked to indicate which category they thought was important to discriminate the representations and which concepts, if any, were missing. These were presented as free-text questions.

The responses to the free-text questions were recorded in tabular form and categorised according to their information content. Particular attention was paid to specific text passages to generate a meaningful structure~\cite{patton_qualitative_2014}. The structured approach made it possible to identify and analyse the frequencies and correlations of the topics and impressions about the details of the experts' answers.  

\subsection{Statistical analysis}\label{subsec2.5}

We calculated the mean and median values across design, functions, tasks, and cross-concepts for each representation. Then, we narrowed our focus to the 16 categories and analysed the differences between the four representations. The coefficient of variation was calculated for each rating item. This represents the standard deviation in relation to the mean value and is dimensionless. If the coefficient of variation is less than or equal to $0.5$, we can say that at least 50\% of the experts provided ratings close to the mean value~\cite{zinn_identifying_2001, von_der_gracht_consensus_2012}. For each of the 140 rating items within the representations, we checked whether there was at least 50\% agreement. If an item rating of a representation fell below this value, this item would not be considered further for statistical analysis, because this indicates that there was substantial disagreement among the raters.

Next, the Levene test was used to test the homogeneity of the variances between the representations. The Friedman test was then performed to determine whether there were significant differences in the mean rating values between the representations. For each rater, the ratings for each representation were converted into ranks, with the lowest rating ranked first. For equal ratings, the ranks were assigned as the average of the positions of the tied ratings. The ranks were summed for each representation to obtain the rank sum for each condition. A significant Friedman test result indicates that there are differences between the representations~\cite{field_discovering_2012}.

Raters evaluated four different representations. Each representation was rated by the same raters in different conditions, such as quantum measurement, superposition and probability.The ratings were made on a scale from 1 to 5, and since the same raters judged each representation in each condition, the ratings were related (dependent).

Finally, the post hoc Wilcoxon signed rank test with Bonferroni correction was used to identify specific differences between the representations~\cite{field_discovering_2012}. This methodological approach enabled a precise and differentiated analysis of the data, providing deeper insights into the variance and significance within the representations studied. In order to identify the impact of the significant values among the representations, Cohen's effect size \textit{d} was calculated~\cite{cohen_statistical_1988}. 

We also calculated the correlations between the overarching categories, design, functions, tasks, and cross-concepts, with Spearman’s correlation coefficient and used this to refine our analysis of the categories. All statistical analyses were performed using R (version 4.4.0, R Core Team, 2024). The R code used for the analysis is available on request.

\section{Results}\label{sec3}
\subsection{Variations in expert ratings}\label{sec3.1}

We calculated the coefficient of variation to determine the level of agreement between raters for each rating item. All coefficients of variation (\autoref{secA2}) were at or below 0.5, which is usually accepted as the threshold for adequate internal agreement~\cite{von_der_gracht_consensus_2012, zinn_identifying_2001}, with the exception of those for the items in the categories \textit{actions/steps}, and \textit{complementarity}, which were all above 0.5. In these cases, no statement could be made due to disagreement among the experts.

A similar disagreement occurred in the rating of the entanglement concept, which was therefore excluded from the \textit{concepts} category. The ratings for global and relative phase visualisation were made separately for the category \textit{phase}. Due to the disagreement in the assessment of the global phase, this was also not taken into account. 

For \textit{understanding difficulties} in item statement B, generability for the Qake representation, and \textit{effort in explanation} for the Quantum Bead representation, the level of agreement was  49\% agreement. Further consideration of the category understanding difficulties was limited to statements A, C, and D, with statement B excluded due to discrepancies between the experts. The same procedure was applied to colour for representation, generability for the Qake representation, and effort in explanation  representation Quantum Bead. If two or more representations within a category had a coefficient of variation greater than or equal to 0.5, they were excluded completely.

\subsection{Considered categories after agreement}\label{sec3.2}

\autoref{Tabelle_meanvalue} shows the means for all categories and the significance within them. The mean values were inverted for specific categories (\textit{understanding difficulties, colour and effort in explanation}) to provide an overview of which categories were rated higher and therefore have a positive impact on learning. For answering RQ1 and RQ2, details are provided in Sections \ref{sec3.2.1} - \ref{sec3.2.4}.

\begin{table}[ht]
\centering
\caption{Mean value of expert ratings for each category}
\begin{tabularx}{\textwidth}{l *{4}{>{\centering\arraybackslash}X} c}
\toprule
\multicolumn{6}{c}{} \\
\multicolumn{6}{c}{Design} \\
\cmidrule(r){1-1} \cmidrule(lr){2-6}
Category & Quantum Bead & Circle Notation & Bloch sphere & Qake & p-value \\
\cmidrule(r){1-1} \cmidrule(lr){2-6}
Salience  & $3.33 \pm 1.03$ & $2.88 \pm 0.93$ & $2.79 \pm 0.90$ & $3.11 \pm 0.88$ & *** \\
Dimension & $3.40 \pm 1.10$  & $3.97 \pm 1.10$  & $3.66 \pm 1.19$ &  $3.99 \pm 1.07$  & ***\\
Understanding difficulties\footnotemark[1] 
          & $3.20 \pm 1.32$  & $4.24 \pm 0.90$  & $3.61 \pm 1.14$ &  $3.82 \pm 1.09$  & *** \\
colour\footnotemark[2]  
          & - & $4.08 \pm 1.12$ & $4.90 \pm 0.31$ & $1.29 \pm 0.46$ & *** \\
\bottomrule
\multicolumn{6}{c}{} \\
\multicolumn{6}{c}{Functions} \\
\cmidrule(r){1-1} \cmidrule(lr){2-6}
Actions/steps
            & - & - & - & - & - \\
Mathematics & $3.18 \pm 1.29$ & $3.94 \pm 1.00$ & $4.16 \pm 0.60$ & $3.71 \pm 0.92$ & n.s. \\
Contiguity  & $3.41 \pm 1.33$ & $3.74 \pm 1.10$ & $3.16 \pm 1.30$ & $3.50 \pm 0.99$ & n.s. \\
Predictability & $3.82 \pm 0.73$  & $4.32 \pm 0.58$ & $4.05 \pm 0.71$ & $4.11 \pm 0.74$ & * \\
\bottomrule
\multicolumn{6}{c}{} \\
\multicolumn{6}{c}{Tasks/applications} \\
\cmidrule(r){1-1} \cmidrule(lr){2-6}
Concepts & $3.38 \pm 1.16$ & $4.02 \pm 1.07$ & $3.41 \pm 1.16$ & $4.19 \pm 1.13$ & *** \\
\hspace{1.5mm} Quantum measurement & $3.48 \pm 1.08$ & $4.00 \pm 1.18$ & $3.66 \pm 1.06$ & $4.19 \pm 1.12$ & ** \\
\hspace{1.5mm} Superposition & $3.24 \pm 1.26$ & $4.10 \pm 1.00$ & $3.62 \pm 1.12$ & $4.24 \pm 1.16$ & ** \\
\hspace{1.5mm} Probabilistics & $3.43 \pm 1.16$ & $3.95 \pm 1.07$ & $3.33 \pm 1.15$ & $4.14 \pm 1.20$ & * \\
Quantum technologies 
         & $3.88 \pm 1.05$ & $3.87 \pm 1.29$ & $4.17 \pm 0.96$ & $3.92 \pm 1.27$ & n.s. \\
Phase (relative) 
         & $3.12 \pm 1.36$ & $4.53 \pm 0.61$ & $4.56 \pm 0.76$ & $4.21 \pm 0.85$ & *** \\
Amplitude & $3.30 \pm 1.03$ & $4.50 \pm 0.51$ & $3.85 \pm 0.88$ & $4.60 \pm 0.50$ & *** \\
\bottomrule
\multicolumn{6}{c}{} \\
\multicolumn{6}{c}{Cross-concepts} \\
\cmidrule(r){1-1} \cmidrule(lr){2-6}
Generability
   & $2.10 \pm 1.22$  & $3.29 \pm 1.06$  & $2.52 \pm 0.75$& -  & ***\\
Effort in explanation\footnotemark[3] 
& - & $2.81 \pm 1.03$ & $2.10 \pm 0.89$ & $3.15 \pm 1.14$ & ** \\
\bottomrule
\end{tabularx}
\footnotetext{The $p$-value was calculated using the Friedman test to determine the difference between the representations. ***$p < 0.001$, **$p < 0.01$, and *$p < 0.05$. n.s. = not significant}
\footnotetext[1]{5 means less prone to difficulties in learning. 1 means very prone to difficulties in learning}
\footnotetext[2]{A high mean value means less need for representations to visualise concepts with colour. This category is based on the need for colour representation.}
\footnotetext[3]{A high mean value means less "effort in explanation" of the representation. }
\label{Tabelle_meanvalue}
\end{table}

\begin{table}[!ht]
\caption{Supplementary mean values of function categories}
\label{tab:beispieltabelle}
\footnotesize 
\begin{tabularx}{\textwidth}{l*{6}{X}c}
\toprule
\multicolumn{8}{c}{Function} \\
\midrule
Category & A & B & C & D & E & F & p-Value \\ 
\midrule
Redundancy  & \tiny$3.05 \pm 1.00$ & \tiny$3.45 \pm 0.89$ & \tiny$3.15 \pm 1.09$ & \tiny$3.50 \pm 1.32$ & \tiny$3.95 \pm 1.05$ & \tiny$3.20 \pm 1.28$ & ** \\ 
Complementarity & - & - & - & - & - & - & - \\ 
\bottomrule
\end{tabularx}
\footnotetext{Combinations of representations A–F: A = Quantum Bead and Circle Notation, B = Bloch sphere and Quantum Bead, C = Qake and Quantum Bead, D = Bloch sphere and Circle Notation, E = Circle Notation and Qake, F = Qake and Circle Notation}
\end{table}
\newpage

\subsubsection{Results from design categories}\label{sec3.2.1}
Based on the data, we observe that there are significant differences between the representations in the ratings for \textit{salience} ($\chi^2(3) = 21.706$, $p < .001$). The rating for the Quantum Bead was significantly higher than for the Bloch sphere ($d=0.36$) and the Circle Notation ($d=0.38$). In addition, there was a significant difference between Qake and the Circle Notation with $d=0.31$, and between Qake and the Bloch sphere ($d=0.32$). The Quantum Bead (mean: $3.33 \pm 1.03$) was rated the most salient, followed by Qake (mean: $3.11 \pm 0.88$). 

Furthermore, there are significant differences in the category \textit{dimension} ($\chi^2(3) = 17.810$, $p < .001$), particularly between the Quantum Bead and the Circle Notation ($d=0.39$) representations and between Qake and Quantum Bead ($d=0.38$). The experts rated the Circle Notation and Qake as more adequate regarding their spatial dimensionality. There is also a significant positive correlation \(r = 0.41\) between \textit{dimension} and \textit{salience} (95\% confidence interval of 0.31 to 0.50).

We found significant differences between the representations in the generation of misconceptions respectively \textit{understanding difficulties} ($\chi^2(3) = 37.090$, $p < .001$). According to the expert ratings, the Circle Notation tends to cause fewer difficulties than the Quantum Bead, ($p < .001$, $d=0.50$), the Bloch sphere ($p < .034$, $d=0.38$), and the Qake representation ($p < .042)$, $d=0.36$). The experts rated the Bloch sphere and the Quantum Bead as more likely to lead to understanding difficulties than the Circle Notation and Qake.

We also found significant differences between the representations in the category \textit{colour} ($\chi^2(2) = 38.297$, $p < .001$). Significant differences were found between the Bloch sphere and Qake ($p < .001$, $d=0.90$) and between the Circle Notation and the Bloch sphere ($p < .001$, $d=0.69$). There was also a significant difference between Qake and the Circle Notation ($p < .001$, $d=0.86$). These differences all have a high effect size. The Bloch sphere was rated high as colour independent respectively no relevant in colour (mean: $4.90 \pm 0.31$) while the Qake model was rated highly colour dependent to visualise the concepts (mean: $1.29 \pm 0.46$). 

\subsubsection{Results from function categories}\label{sec3.2.2}
The results indicate non-significant differences between the representations in terms of \textit{mathematics} and \textit{contiguity}. Despite the similar rating regarding interaction with \textit{mathematics}, the Circle Notation and Qake were rated significantly higher in regarding whether concepts ($\chi^2(3) = 37.258, p < .0001$) such as quantum measurement, superposition, and probabilistics were appropriately visualised. 

The data indicate a significant difference in \textit{predictability} ($\chi^2(3) = 10.451, p = .015$). Moreover, there is a significant positive correlation \(r = 0.47\) between the ratings for \textit{predictability} and \textit{mathematics} (95\% confidence interval of 0.27 to 0.63).%

\begin{figure}[ht]
    \includegraphics[width=0.65\textwidth]{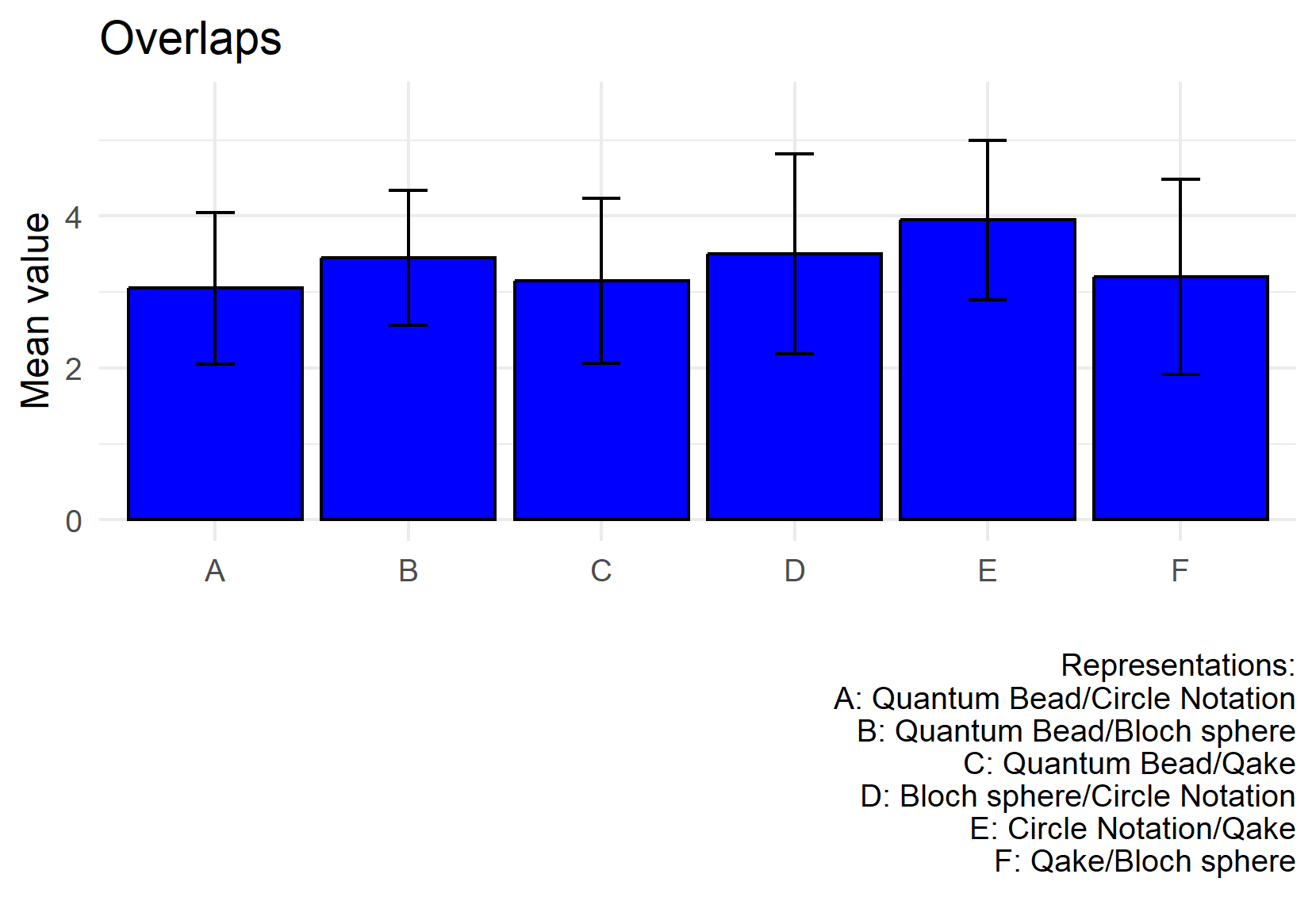}
    \caption{Experts’ ratings of whether representations have the same information content of concepts.}
    \label{Overleaps}
\end{figure}

The experts rating also analysed how representations are used in combination with each other \textit{overlap/redundancy} \autoref{Overleaps}), i.e., how many complementary or redundant information they contain. The category \textit{complementary} was not analysed further due to disagreement among the experts. However, there is a significant difference between particular groups in \textit{overlap/redundancy} ($\chi^2(5) = 16.126, p < .01$). Between groups A and E ($p < .05$) a significant difference (with an effect size of $d=0.64$) could be determined, as wel as between groups B and F ($p=.05$, and $d=0.19$). Between all pairs, no significant difference could be identified. Representations with the same dimensions were assigned the highest rating values for \textit{overlap/redundancy} (B: $3.45 \pm 0.89$; E: $3.95 \pm 1.05$).

\vspace{-2.0\baselineskip}
\subsubsection{Results from task categories}\label{sec3.2.3}
As already mentioned, there was a significant difference between the expert ratings of the suitability of the representations for visualising \textit{concepts} ($\chi^2(3) = 37.258, p < .001$). This concerns the concepts of quantum measurement, superposition, and probabilistic. This difference was found to exist mainly between the Circle Notation and the Bloch sphere ($d = 0.43$), the Quantum Bead and the Circle Notation ($d = 0.42$), Qake and the Bloch sphere ($d = 0.48$), and Qake and the Quantum Bead ($d = 0.51$). Circle Notation (mean: $4.02 \pm 1.07$) and Qake (mean: $3.88 \pm 1.05$) were rated higher on average.

The results indicate significant differences in the representations regarding the visualisation of \textit{phase (relative)} ($\chi^2(3) = 15.528$, $p < .001$). Post hoc tests showed that these differences were mainly between the Quantum Bead and Bloch sphere representations ($p < .01$, $d = 0.64$), and also between the Quantum Bead and Circle Notation ($p < .01$, $d = 0.60$). A high effect size can be assigned to this. The Circle Notation and Qake were rated higher on average.

There was also a significant difference in the visualisation of the \textit{amplitude} ($\chi^2(3) = 29.008$, $p < .001$). The data showed a difference between the Quantum Bead and Circle Notation representations ($p < .01)$, $d = 0.67$) and also a significant difference between the Quantum Bead and Qake ($p < .01$, $d = 0.76$). Significant differences were also found in the comparisons between the Qake representation and the Bloch sphere ($d = 0.66$). Qake (mean: $4.60 \pm 0.52$) and the Circle Notation (mean: $4.50 \pm 0.51$) were rated higher on average. A high effect size can be assigned to this. 

Moreover, ratings for visualisation of the \textit{amplitude} correlate positively with those for the \textit{predictability} category \(r = 0.49\) (95\% confidence interval of 0.30 to 0.64). 

There are no significant differences between the representations concerning the category \textit{quantum technology}. This is likely due to the fact that all representations of a qubit were visualised, thus allowing us to demonstrate the elementary operation of a quantum computer (the X-, Z-, and H-gates). 

\subsubsection{Results from cross-concept categories}\label{sec3.2.4}
The data indicated significant differences between the representations in terms of effort in explanation ($\chi^2(2) = 8.9333$, $p < .01$) specifically between the Circle Notation and the Bloch sphere (d = 0.55, p = .034) and between Qake and the Bloch sphere ($d = 0.56$, $p = .032$).

The expert assessments differed significantly across representations for the category \textit{generability} according to the Friedman test results ($\chi^2(2) = 8.4595$, $ p < .01$). However, the subsequent pairwise comparisons using the Wilcoxon signed-rank test with Bonferroni correction showed no significant differences between the individual representations \((p > .05)\). This suggests that, although there were overall differences in the ratings, none of the specific pairwise differences were statistically significant after the correction for multiple comparisons was applied.

The coefficients of variation for \textit{generability} the Qake representation and \textit{effort in the explanation} for the Quantum Bead exceeded 0.5, indicating significant disagreement among the experts. Due to the lack of consensus reflected in the data, these representations were excluded from consideration in these categories.  
\\

\begin{figure}[ht]
\centering
    \includegraphics[width=0.9\textwidth]{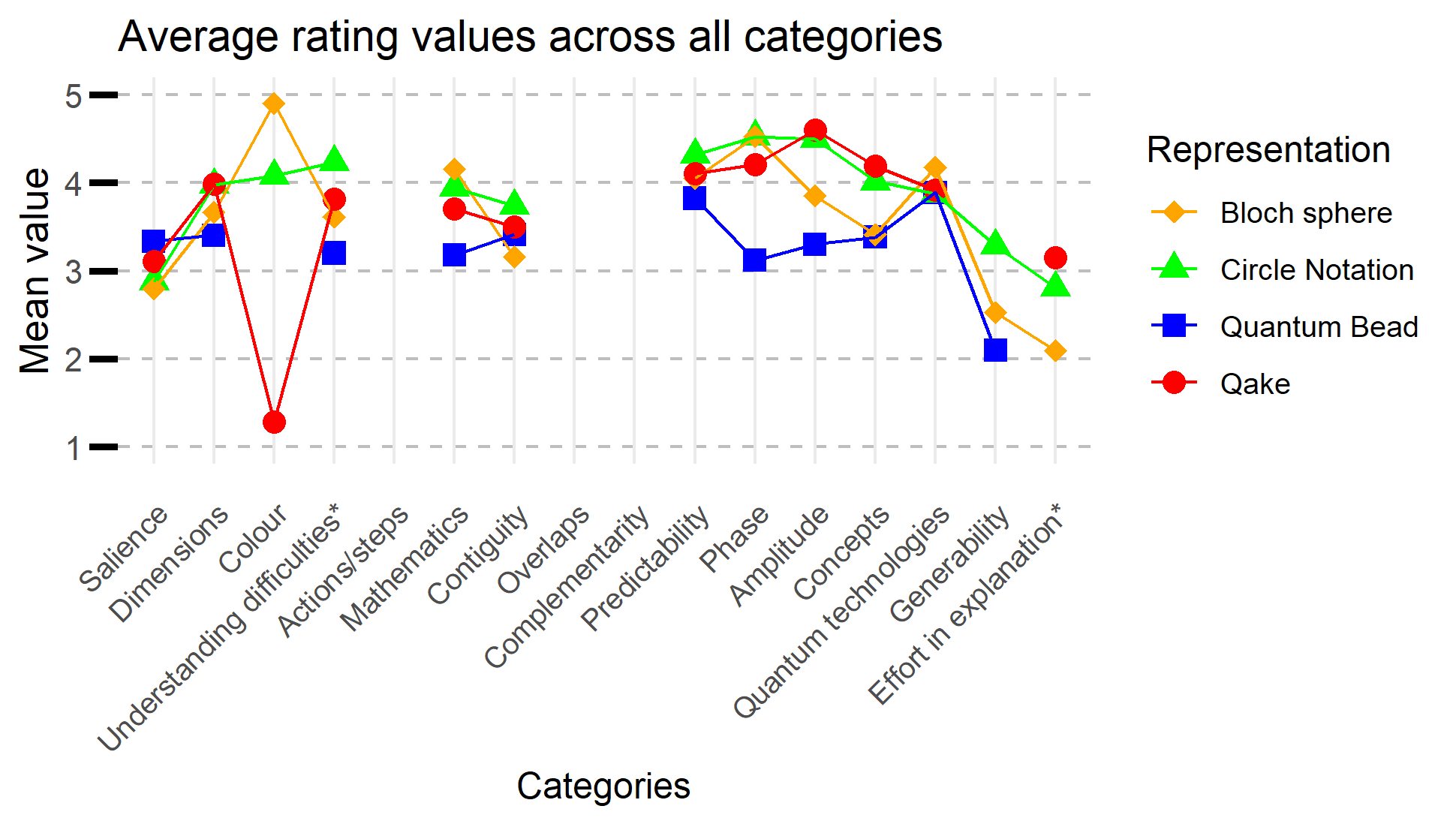}
    \caption{The mean values were inverted for \textit{understanding difficulties}, that is, high values in these items mean that experts associate the representation with a lower tendency to cause understanding difficulties, because we want to give an overview in this table of which categories were rated higher and are, therefore, positive for learning. \textit{Effort in explanations} is also inverted: high values in these cases mean that experts perceive less effort in presenting information with the representation. The categories and representations that were not analysed further due to a lack of agreement are as follows: \textit{actions/steps and complementarity}, \textit{colour} for the Quantum Bead, \textit{generability} for the Cake, and \textit{effort in explanations.} for the Quantum Bead}
    \label{Rating overview of all Categoies (normed)}
\end{figure}

In the following, we present an overview of the results and indicate how they can play a role in the development of new qubit representations. \autoref{Rating overview of all Categoies (normed)} visualises how the four representations differ in their profiles across categories. The categories \textit{overlap/redundancy} and \textit{complementarity} are not included as they do not refer to individual representations but to combinations of them.

It can be seen that representations were rather similar regarding the categories \textit{contiguity, mathematics, and quantum technologies}. However, the experts perceived considerable differences between the representations in the following categories: \textit{phase, amplitude, concepts,} and \textit{understanding difficulties}. Moderate differences were observed for \textit{salience, dimension}, and \textit{predictability}. \autoref{tab:effektgroessen} shows the effect sizes of the differences within the representations. These can be helpful when deciding on the choice and creation of new representations, in which a categorisation can be made beforehand according to similarities and average, strong differences in the aspects of representations. It is essential to plan how these are used in their tasks/applications. The mean value was inverted for \textit{understanding difficulties} to provide an overview of which categories were rated higher and thus have a positive impact on learning. \textit{Colour} is intended to show how relevant the colour component is for understanding the concepts and represents a limitation of the representation, which is why it is inverted.

\captionsetup[table]{skip=10pt}
\begin{table}[!ht]
    \centering
    \caption{Overview of effect size for each category}
    \label{tab:effektgroessen}
    \begin{tabular}{>{\raggedright\arraybackslash}p{5cm} >{\raggedright\arraybackslash}p{6cm}}
        \toprule
        \textbf{Category} & \textbf{Effect sizes} \\
        \midrule
        \rowcolor{lime!60} \textbf{Moderate to high effect} & \\
        Amplitude & $0.66 \leq d \leq 0.76$ \\
        Phase \tiny*relative & $0.60 \leq d \leq 0.64$ \\
        Understanding difficulties & $0.40 \leq d \leq 0.52$ \\
        Overlap/redundancy & $d=0.64$\\
        \midrule
        \rowcolor{lime!30} \textbf{Moderate effect} & \\
        Concepts & $0.42 \leq d \leq 0.51$ \\
        Dimension &   $d \leq 0.39$ \\
        Salience & $0.31 \leq d \leq 0.38$ \\
        \midrule
        \rowcolor{pink!40} \textbf{No difference} & \\
        Mathematics & \\
        Contiguity & \\
        Predictability & \\
        Quantum technologies & \\
        \bottomrule
    \end{tabular}
\footnotetext{Only effect sizes for which the experts were in agreement for all representations are shown. Thus, \textit{colour, generability and effort in explanations} are not listed.}
\end{table}

\newpage
\subsubsection{Results of the free-text questions}\label{sec3.2.5}
The experts were also asked to indicate which \textbf{category} they thought was important to differentiate the representations and which concepts, if any, were missing. 

An expert noted that the \textit{salience} category is difficult to assess, as it is hard to fulfil the conditions for salience via the rating.

The \textit{actions/steps} category is also difficult to quantify. This is also reflected in the wide spread of rater responses in this category, which is still meaningful for categorising the representations, but should be recorded differently. One possibility would be to record the individual steps qualitatively via what is spoken or written and to quantify them via defined sections in “steps”. 

Another expert, felt that the \textit{effort in explanation} category is the most important from an educational perspective. This proves to be a good categorisation of the representation for effective and target group-oriented use.

Some experts suggested including additional categories for correctness, enjoyment in learning, the possible construction of real models for visualisation, and interactive software or videos. 

The following additional \textbf{concepts} were suggested: mixed states, partial traces, errors, types of entanglement, entanglement entropy, systems with more than two qubits, C-X (controlled-X) gates for two qubits in all representations, as this is essential for quantum computing, and visualisation of multiple qubits and algorithms. 

\section{Discussion}\label{sec4}

The expert rating process had two goals: to identify features of qubit representations that may support learning and to identify factors that need to be considered when developing new qubit representations. 

\subsection{Interpretation of findings}\label{sec4.1}

The data show that, according to the experts, the Circle Notation and Qake are especially well suited for visualising concepts such as phase and amplitude. They provide clarification of the theoretical relationships between phase and amplitude, which play a central role in the precise description of a quantum state.

In general, the aim in learning environments is to reduce difficulties that can be caused by misconceptions in visual-graphical representations. More attention needs to be paid to shape, dimension, and preconceptions to avoid misunderstandings ~\cite{ubben_gestalt_2021, singh_review_2015, heo_learning_2020}. The exclusion of one statement (B) has shown that it is difficult to judge on the basis of statements whether visual-graphical representations can cause or reinforce misconceptions or difficulties.

The remaining statements allow us to identify the differences in perception between the raters regarding the potential for triggering difficulties or false concepts in learners, with a medium to high effect size, which is why we think it is important to consider difficulties with regard to misconceptions.
 
According to the ratings, the qubit representations differ in \textit{salience}, which may play a key role in learning. The correlation of the ratings in the salience category with those in the \textit{dimension} category suggests that more \textit{salient} representations like the Quantum Bead also have more appropriate dimensionality (2D or 3D) for understanding the concept. It is possible that this correlation is due to other factors, or that the variety of representations employed is insufficient to fully recognise this effect. Particularly highly salient representations show a strong colour component (e.g. the Quantum Bead). Brightly coloured representations such as the Quantum Bead are beneficial for learning if they are used to attract attention to relevant components~\cite{mayer_signaling_2014}.  

Moreover, the expert rating results indicate that the dimensionality of a representation could play a role in \textit{understanding difficulties}. In consideration of the theory and the findings ~\cite{castro-alonso_sex_2019, heo_learning_2020, saha_he_2016}, it has already been shown that spatial dimensionality can lead to difficulties in understanding. Overall, however, the expert ratings only refer to a small, selected number of statements regarding difficulties or misconceptions, which could potentially be caused by learners being overwhelmed by or misinterpreting the representations’ features.

There are no significant differences between the representations regarding the \textit{function} categories for qubit representations, except for \textit{predictability} and \textit{overlap/redundancy}. This is likely because most categories were considered elementary for qubit representations or because the selected representations may be similar in these respects.

The expert rating has shown that the individual qubit representations have different feature profiles, which is reflected in the different strengths and weaknesses of the representations. None of the representations could be found to be clearly superior or inferior in all categories. In other contexts, the use of several  representations, also known as multiple representations, has proven to be useful (see Section \ref{subsec1.3}). The following quotes from experts in the free-text responses illustrate this:

\begin{description}
    \item\textit{"[...] But there should never be the "one representation", since more visual models like the qubit cakes will always be easier to digest at first, while models like the bloch-sphere help understand more complex topics.” (Expert no. 16)}
\end{description} 
\begin{description}
    \item\textit{“The Circle Notation and the quantum cake model purely represent the mathematical tool of expressing a qubit (or several qubits). Hence they are mostly helpful to students who are struggling with the basic math. The other two representations (Quantum Bead and Bloch sphere) on the other hand are somewhat more advanced, because they try to represent the fact that a qubit can be visualized in 3-dimensional space. [...]” (Expert no. 3)}
\end{description}

This emphasises how important it is to know features/aspects of qubit representations in order to use them in a targeted and learning-promoting way.

\subsection{Further research}\label{sec4.2}

An attempt has been made to draw conclusions for learners from the experts’ assessments and current learning research, but further studies are needed to verify the results and obtain a more detailed perspective. \autoref{tab_furtherres} provides an overview of expert opinions on the individual features of the representations using the mean values. This helps to raise questions for follow-up studies, such as which of the features that differ between the representations according to the experts’ ratings are relevant to learning. The scaling from low to high was chosen based on whether there were significant differences in the data from the tests reported in this study and the corresponding orientation by mean value. If there were no significant differences between the representations, the same category was chosen for them. 
In future studies that investigate the influence of single features of representations on learning, as many aspects as possible should be controlled to study the effect of a particular aspect. Maintaining this balance is a challenge. The gains in learning provided by representations could be measured. In any case, a follow-up study is required to capture the effect of the representations on learners’ achievements. It is, therefore, necessary to investigate whether the identified features of qubit representations, which differ from expert opinion, can be transferred to the perspective of the learners and examined for their potential to facilitate learning. For instance, it could be investigated whether certain representations are more likely to cause difficulties than others. It would also be interesting to investigate the relationship between predictability and amplitude visualisation in more detail. 
Are there effects on learning when the amplitude is visualised more strongly in the representations?

\textbf{Draft of a study:} In \autoref{tab_furtherres}, the differences and similarities between the representations are shown. One possible study would be to first analyse two similar representations that nevertheless have substantial differences in a certain category, such as the Quantum Bead and the Bloch sphere in the salience category. The challenge, however, is to reduce or avoid compensation effects or effects that are stronger due to other categories than in the other representation (in this case, phase and amplitude for the Bloch sphere). Both representations could be delivered via a learning unit for a specific concept, for example, for superposition. Eye tracking could be used to analyse the learners’ gaze behaviour and determine the duration of fixation. Pre- and post-tests could be used to compare the learning gain with the representations and thus determine how high the learning gain was when using the representations. Many other studies are possible.

\definecolor{verylow}{rgb}{1.0, 0.0, 0.0}
\definecolor{low}{rgb}{1.0, 0.4, 0.4}
\definecolor{middle}{rgb}{1.0, 0.8, 0}
\definecolor{high}{rgb}{0.6, 0.8, 0.2}
\definecolor{grey}{rgb}{0.7, 0.7, 0.7}

\begin{table}[!ht]
\centering
\caption{Scaling of the different categories in relation to representations}
\begin{tabular}{lcccc}
\toprule
\textbf{Categories} & Quantum Bead & Circle Notation & Bloch sphere & Qake \\
\midrule
Salience & \cellcolor{high}high & \cellcolor{low}low & \cellcolor{low}low & \cellcolor{middle}middle \\
Dimension & \cellcolor{middle}middle & \cellcolor{high}high & \cellcolor{middle}middle & \cellcolor{high}high \\
Understanding difficulties\footnotemark[1] & \cellcolor{middle}middle & \cellcolor{high}high & \cellcolor{middle}middle & \cellcolor{middle}middle \\
color\footnotemark[2] & \cellcolor{grey}\textcolor{black}{no information} & \cellcolor{middle}middle & \cellcolor{high}high & \cellcolor{low}low \\
Predictability & \cellcolor{middle}middle & \cellcolor{high}high & \cellcolor{middle}middle & \cellcolor{middle}middle \\
Concepts & \cellcolor{middle}middle & \cellcolor{high}high & \cellcolor{middle}middle & \cellcolor{high}high \\
Phase (relative) & \cellcolor{middle}middle & \cellcolor{high}high & \cellcolor{high}high & \cellcolor{middle}middle \\
Amplitude & \cellcolor{low}low & \cellcolor{high}high & \cellcolor{middle}middle & \cellcolor{high}high \\
Generability & \cellcolor{middle}middle & \cellcolor{middle}middle & \cellcolor{middle}middle & \cellcolor{grey}\textcolor{black}{no information} \\
Effort in explanations & \cellcolor{grey}\textcolor{black}{no information} & \cellcolor{high}high & \cellcolor{low}low & \cellcolor{high}high \\
\bottomrule
\end{tabular}
\label{tab_furtherres}

\footnotetext[1]{"High" means less prone to difficulties; "low" means very prone to difficulties}
\footnotetext[2]{"High" means that colour are not necessary to visualise concepts in QP, which is positive for representation.}
\footnotetext[3]{"High" means less effort is needed to explain how a representation presents information; “low” means more effort is needed.}
\end{table}

As mentioned, MER can be used in different ways. When there are overlaps or redundancies between representations, these can be useful for learning concepts in QP and QT. In favour of the use of multiple representations: overlaps in representations of the same spatial dimensionality were rated higher on average by the experts. Further research is needed to investigate the learning potential of the use of multiple representations and, in particular, which combinations are conducive to learning. We know from MER research that the use of multiple representations can fulfil various functions that can promote learning with representations~\cite{ainsworth_deft_2006}. They can be analysed for different functions. Specifically, there are three key functions that MERs can perform (even simultaneously) to support the learning process:
\begin{enumerate}
    \item supporting complementary processes
    \item enabling representations to constrain each other
    \item developing of deeper understanding
\end{enumerate}

As part of a study, it could be examined to what extent and with combinations the use of MER for qubit representations makes sense. The combinations A–F from the rating can be used for this purpose. The study could contrast learning using a combination of representations with learning using individual representations. 

\subsection{Limitations}\label{sec4.3}

The expert ratings provide us with a categorisation based on people with a wealth of experience, but there are also some limitations, such as the small sample size and the specific selection of representations. The experts had a limited view of the representations as they only had access to the contents of the cheat sheet and, where applicable, material from the references. It is therefore possible that the strength of the representations was limited.

In addition, the categories that were not further analysed due to a lack of agreement are not necessarily unsuitable, including \textit{complementarity} and \textit{actions/steps}. With a larger sample, it could be valuable to analyse these further.

Due to the lack of consensus among the raters, items in certain categories or even entire item sets, and thus the entire category, were excluded, so that the expert ratings only offer a limited view of the features of visual-graphical qubit representations. The concepts were exclusively focused on the consideration of single-qubit cases. Multi-qubit systems were not analysed.

\section{Conclusion}\label{sec5}

The expert ratings provide key insights into the features of qubit representations that facilitate learning in QP and highlight important considerations for developing new, more effective representations in QP education. Choosing appropriate representations is challenging. Leveraging the category system based on the DeFT framework could help make visual representations more effective and optimise their use in different educational contexts.

The expert ratings in this study suggest that representations like the Circle Notation or Qake could be more effective in teaching fundamental QP concepts. The Circle Notation was also rated highly in terms of phase and amplitude and the experts attached particular importance to representations that clearly visualise these aspects. In addition, representations that explicitly visualise the relative phase with an arrow or a line, such as the Circle Notation and the Bloch sphere, received higher ratings. Although colour can enhance salience for specific purposes, it is not deemed essential for conveying the core concepts. Misconceptions can arise from different sources, and careful consideration of these factors is necessary when developing educational materials.

Key factors in developing new qubit representations include minimising difficulties, particularly through clear visualisation of the (relative) phase and amplitude. Our findings indicate that the Circle Notation and the Qake model are particularly effective in visualising quantum concepts and are considered easier for learners to understand. To prevent misunderstandings, especially with less intuitive representations like the Quantum Bead and the Bloch sphere, a more detailed introduction or explanation may be required. This suggests that the Circle Notation or a pie chart model, like Qake, may be more suitable or efficient for teaching the fundamental concepts of QP and QT (see \autoref{Rating overview of all Categoies (normed)} and \autoref{tab:effektgroessen}).

Finally, representations such as the Bloch sphere, cannot be used for all concepts (e.g. entanglement), so it must be used in combination with other qubit representations when teaching QT. According to experts, they can certainly be used for educational purposes and may even be capable of supporting a transition to the use of MER to fill this “gap”~\cite{ainsworth_deft_2006, rau_conditions_2017}. This enables a comprehensive representation of quantum concepts, particularly those that are fundamental to QT.

These findings have important implications for the development of future teaching materials and could significantly enhance the teaching of QP. Future research should focus on refining these representations for different educational levels and contexts, exploring their effectiveness across diverse learning environments, and integrating additional visual elements to further aid comprehension.

Moreover, this study is not without limitations, including its relatively small sample size, limited range of representations, and exclusive focus on single-qubit cases. Thus, although the findings are promising, further research is needed to confirm the results across broader settings and with more varied representations. Future studies should also investigate multi-qubit systems to provide a more comprehensive understanding of effective representations in QP education.

\section*{List of Abbreviations}
\begin{itemize}[left=0pt]
    \item[QP] Quantum physics
    \item[QT] Quantum technology
    \item[MER] Multiple external representations
    \item[DeFT] Design, Functions and Tasks 
\end{itemize}

\backmatter

\section*{Declarations}

\textbf{Ethical issues and agreeing to participate}

Ethical review and approval was not required for the study involving human subjects in accordance with local legislation and institutional requirements. The study involved data collection using an online questionnaire. Participation was voluntary and anonymous. Participants had the option of giving their name voluntarily. Participants were informed that their answers would be kept confidential and would only be used for research purposes. Data collected included the location of each participant’s research institution, their field of research, and their years of teaching experience. By voluntarily completing the online questionnaire, participants confirmed their participation and consented to the use of their anonymised data for research purposes and publication.

\textbf{Publication consensus}
Not applicable.

\textbf{Competing interests} 
The authors declare that they have no competing interests.

\textbf{Financial Interests}
The authors have no relevant financial or non-financial interests to disclose.

\textbf{Acknowledgements} 
{We would like to thank Tatjana Wilk, General Manager of the Munich Center for Quantum Science and Technology (MCQST), and Silke St\"ahler-Schöpf, Head of the student laboratory PhotonLab at the Max Planck Institute of Quantum Optics, for their valuable contributions to this article. We are also grateful to all participants who contributed to this study.}

\textbf{Funding}
This research is part of the Munich Quantum Valley (MQV), which is supported by the Bavarian State Government.

\textbf{Authors' contributions}
{L.Q., St.K., J.K., S.Ma. designed the study, L.Q., St.K., developed the questionnaires and collected the data, L.Q., St.K., analysed the data, L.Q. wrote the first draft of the manuscript, L.Q., St.K., J.K., S.Ma., E.R., S.M. reviewed and edited the manuscript. St.K. supervised the study. All authors have read and agreed to the submitted version of the manuscript.}

\textbf{Data availability} 
All data and materials are available upon request.

\bigskip

\newpage

\begin{appendices}

\section{Medians}\label{secA1}

\autoref{tab6} and \autoref{tab7} provide information about the median rating levels.

\begin{table}[ht]
\centering
\caption{Median expert rating for each category}\label{tab6}
\begin{tabularx}{\textwidth}{l *{4}{>{\centering\arraybackslash}X} c}
\toprule
\multicolumn{6}{c}{} \\
\multicolumn{6}{c}{Design} \\
\cmidrule(r){1-1} \cmidrule(lr){2-6}
Category & Quantum Bead & Circle Notation & Bloch sphere & Qake & p-value \\
\midrule
Salience  & $3.0$ & $3.0$ & $3.0$ & $3.0$ & *** \\
Dimension & $4.0$  & $4.0$  & $4.0$ &  $4.0$  & ***\\
Understand difficulties\footnotemark[1] 
          & $4.0$  & $3.0$  & $4.0$ &  $3.0$  & *** \\
colour 
          & - & $4.0$ & $5.0$ & $1.0$ & *** \\
\bottomrule
\multicolumn{6}{c}{} \\
\multicolumn{6}{c}{Function} \\
\cmidrule(r){1-1} \cmidrule(lr){2-6}
Actions/steps
            & - & - & - & - & - \\
Mathematics & $3.0$ & $4.0$ & $4.0$ & $4.0$ & n.s. \\
Contiguity  & $4.0$ & $4.0$ & $3.0$ & $4.0$ & n.s. \\
Predictability & $4.0$  & $4.0$ & $4.0$ & $4.0$ & * \\
\bottomrule
\multicolumn{6}{c}{} \\
\multicolumn{6}{c}{Tasks/applications} \\
\cmidrule(r){1-1} \cmidrule(lr){2-6}
Concepts & $3.0$ & $4.0$ & $3.0$ & $5.0$ & *** \\
\hspace{1.5mm} Quantum measurement & $3.0$ & $4.0$ & $4.0$ & $5.0$ & ** \\
\hspace{1.5mm} Superposition & $3.0$ & $4.0$ & $4.0$ & $5.0$ & ** \\
\hspace{1.5mm} Probabilistics & $4.0$ & $4.0$ & $3.0$ & $5.0$ & * \\
Quantum technologies 
         & $4.0$ & $4.0$ & $4.0$ & $4.0$ & n.s. \\
Phase\footnotemark[2] 
         & $4.0$ & $5.0$ & $5.0$ & $4.0$ & *** \\
Amplitude & $3.5$ & $4.5$ & $4.0$ & $5.0$ & *** \\
\bottomrule
\multicolumn{6}{c}{} \\
\multicolumn{6}{c}{Cross-concepts} \\
\cmidrule(r){1-1} \cmidrule(lr){2-6}
Generability
   & $2.0$ & $3.0$ & $2.0$ & -  & ***\\
Effort in explanation\footnotemark[3] 
& - & $3.0$ & $2.0$ & $3.0$ & ** \\
\bottomrule
\end{tabularx}
\footnotetext{The p-value was calculated using the Friedman test to determine the difference between the representations. ***$p < 0.001$, **$p < 0.01$ and *$p < 0.05$. n.s. = not significant}
\footnotetext[1]{Inverted: 5 means very prone to difficulties, 1 means less prone to difficulties}
\footnotetext[2]{Relative Phase}
\footnotetext[3]{Inverted}
\end{table}

\begin{table}[!ht]
\centering
\caption{Supplementary medians of function categories}\label{tab7}
\begin{tabularx}{\textwidth}{l *{6}{>{\centering\arraybackslash}X} c}
\toprule
\multicolumn{8}{c}{Function} \\
\midrule
Category & A & B & C & D  & E & F & p-value \\ 
\midrule
Redundancy  & $3.0$  & $3.5$  & $3.0$ & $4.0$  & $4.0$  & $3.0$ & ** \\ 
Complementaitry & -  & - & - & - & - & - & - \\ 
\bottomrule
\end{tabularx}
\footnotetext{Combinations of representations A–F: A = Quantum Bead und Circle Notation, B = Bloch Sphere and Quantum Bead, C = Qake and Quantum Bead, D = Bloch Sphere and Circle Notation, E = Circle Notation and Qake, F = Qake and Circle Notation}
\end{table}
\newpage

\section{Coefficient of variation}\label{secA2}
\vspace{-1.5\baselineskip}
\begin{table}[!ht]
    \centering
    \caption{Coefficient of variation for the total items of the categories}
    \begin{tabular}{lcccc}
        \hline
        Item & Quantum Bead & Circle Notation & Bloch sphere & Qake \\
        \hline
        Salience\_QS & 0.27 & 0.36 & 0.27 & 0.25 \\
        Salience\_QM & 0.32 & 0.35 & 0.30 & 0.29 \\
        Salience\_Sp & 0.29 & 0.32 & 0.31 & 0.27 \\
        Salience\_E & 0.37 & 0.27 & 0.40 & 0.31 \\
        Dimension\_QS & 0.29 & 0.29 & 0.21 & 0.21 \\
        Dimension\_QM & 0.21 & 0.23 & 0.27 & 0.25 \\
        Dimension\_Sp & 0.27 & 0.19 & 0.27 & 0.20 \\
        Dimension\_E & 0.50 & 0.38 & 0.50 & 0.35 \\
        Actions\_steps\_QS & 0.52 & 0.39 & 0.34 & 0.38 \\
        Actions\_steps\_QM & 0.53 & 0.70 & 0.63 & 0.48 \\
        Actions\_steps\_Sp & 0.56 & 0.52 & 0.53 & 0.48 \\
        Actions\_steps\_E & 0.59 & 0.55 & 0.66 & 0.57 \\
        Understanding\_difficulties\_A & 0.19 & 0.45 & 0.35 & 0.49 \\
        Understanding\_difficulties\_B & 0.51 & 0.47 & 0.41 & 0.51 \\
        Understanding\_difficulties\_C & 0.42 & 0.44 & 0.42 & 0.43 \\
        Understanding\_difficulties\_D & 0.28 & 0.29 & 0.29 & 0.39 \\
        colour & 0.55 & 0.28 & 0.06 & 0.47 \\
        Mathematics & 0.40 & 0.25 & 0.14 & 0.25 \\
        Contiguity & 0.39& 0.29 & 0.41 & 0.28 \\
        Predictability & 0.19 & 0.13 & 0.17 & 0.18 \\
        Phase\_Gl & 0.64 & 0.36 & 0.81 & 0.32 \\
        Phase\_Rl & 0.43 & 0.13 & 0.13 & 0.21 \\
        Amplitude & 0.31 & 0.11 & 0.23 & 0.11 \\
        Concept\_QM & 0.31 & 0.30 & 0.29 & 0.27 \\
        Concept\_Sp & 0.39 & 0.24 & 0.31 & 0.27 \\
        Concept\_E & 0.56 & 0.51 & 0.44 & 0.45 \\
        Concept\_prob & 0.34 & 0.27 & 0.35 & 0.29 \\
        QT\_H-Gate & 0.27 & 0.36 & 0.26 & 0.36 \\
        QT\_X-Gate & 0.23 & 0.31 & 0.18 & 0.26 \\
        QT\_Z-Gate & 0.29 & 0.33 & 0.24 & 0.34 \\
        Generability & 0.32 & 0.39 & 0.22 & 0.51 \\
        Effort\_in\_explanation & 0.51 & 0.37 & 0.42 & 0.36 \\
        \hline
    \end{tabular}
 \footnotetext{Any item with a coefficient of variation of 0.5 or higher was excluded from further analysis.}
    \label{Tabelle_VC}
\end{table}

\begin{table}[!ht]
\centering
\caption{Coefficient of variation for complementarity and overlap/redundancy in combination of representations. We structure them from A to F.}
\label{tab:beispieltabelle2}
\begin{tabularx}{\textwidth}{l *{6}{>{\centering\arraybackslash}X}}
\toprule
Item & A & B & C & D  & E & F \\ 
\midrule
Redundancy  & 0.33  & 0.26  & 0.35 & 0.38  & 0.27  & 0.40  \\ 
Complementarity & 0.46  & 0.51 & 0.48 & 0.53 & 0.48 & 0.50 \\ 
\bottomrule
\end{tabularx}
\footnotetext{Combinations of representations A–F: A = Quantum Bead und Circle Notation, B = Bloch sphere and Quantum Bead, C = Qake and Quantum Bead, D = Bloch sphere and Circle Notation, E = Circle Notation and Qake, F = Qake and Circle Notation}
\end{table}

\section{Mean rating value (graphs)}\label{secA3}

\begin{figure}[H]
    \begin{subfigure}[b]{0.49\textwidth}
        \includegraphics[width=\textwidth]{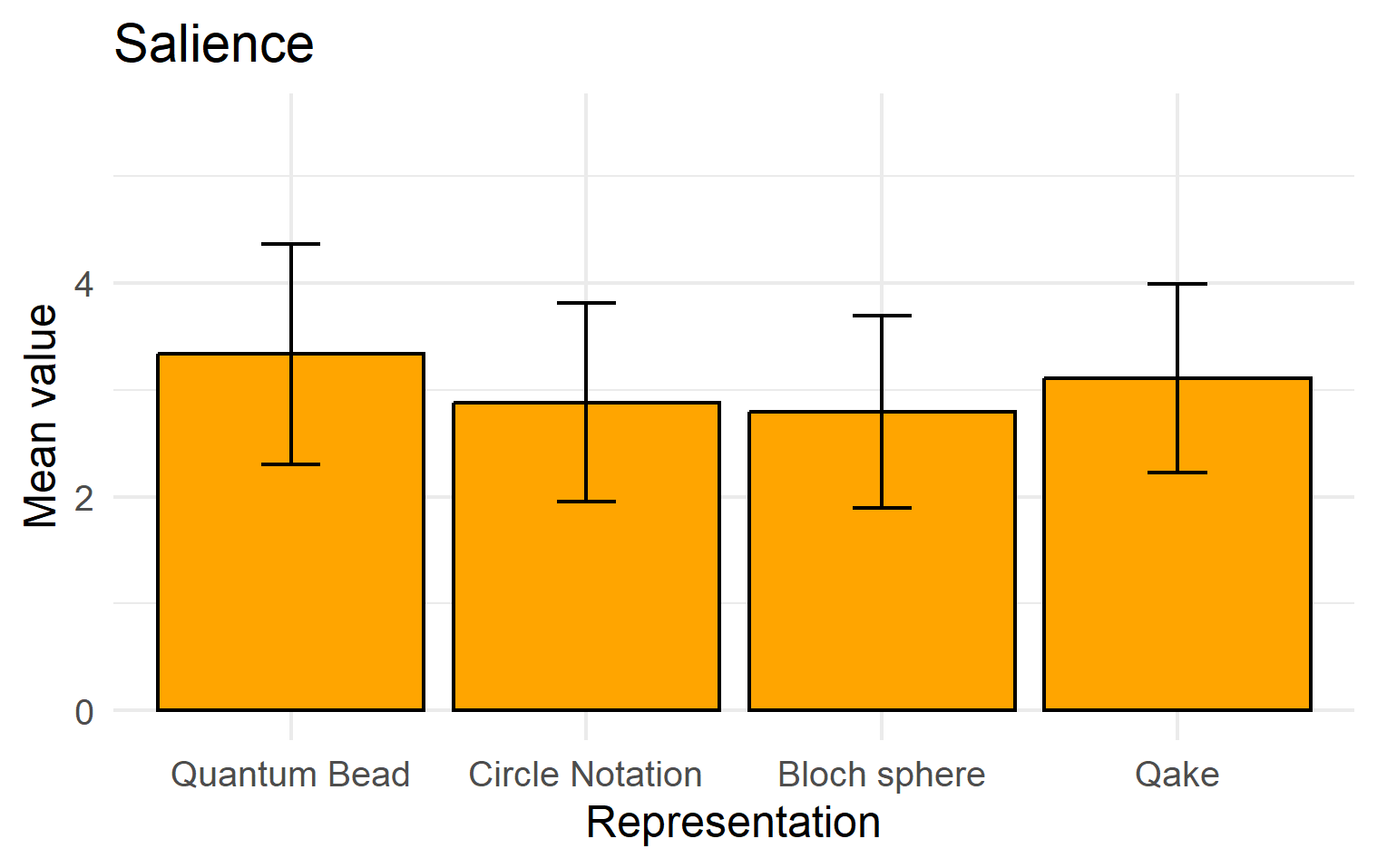}
        \caption{Salience}
        \label{fig:mein_bild_1}
    \end{subfigure}
    \hfill
    \begin{subfigure}[b]{0.49\textwidth}
        \includegraphics[width=\textwidth]{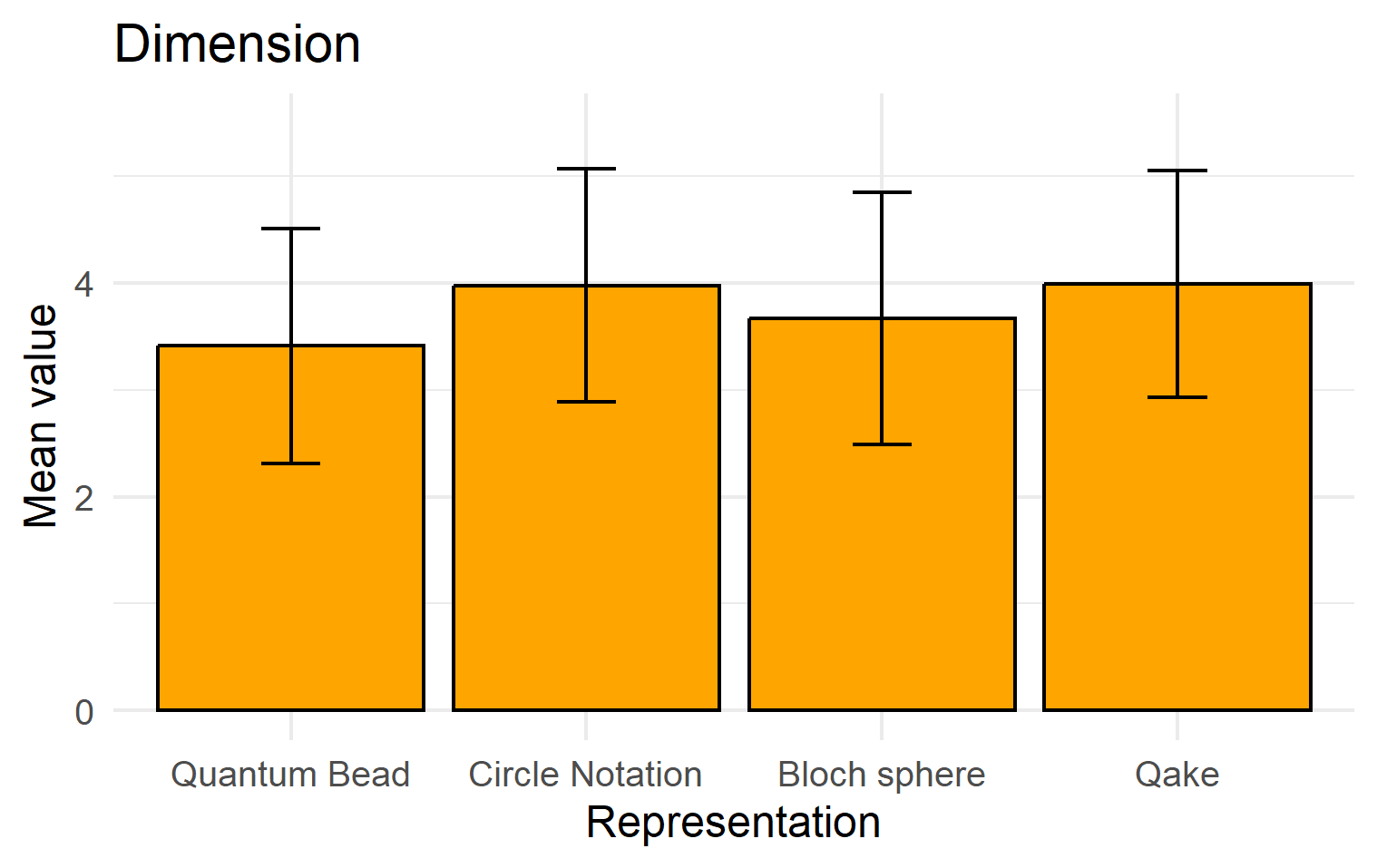}
        \caption{Dimension}
        \label{fig:mein_bild_2}
    \end{subfigure}
    \hfill
    \begin{subfigure}[b]{0.49\textwidth}
        \includegraphics[width=\textwidth]{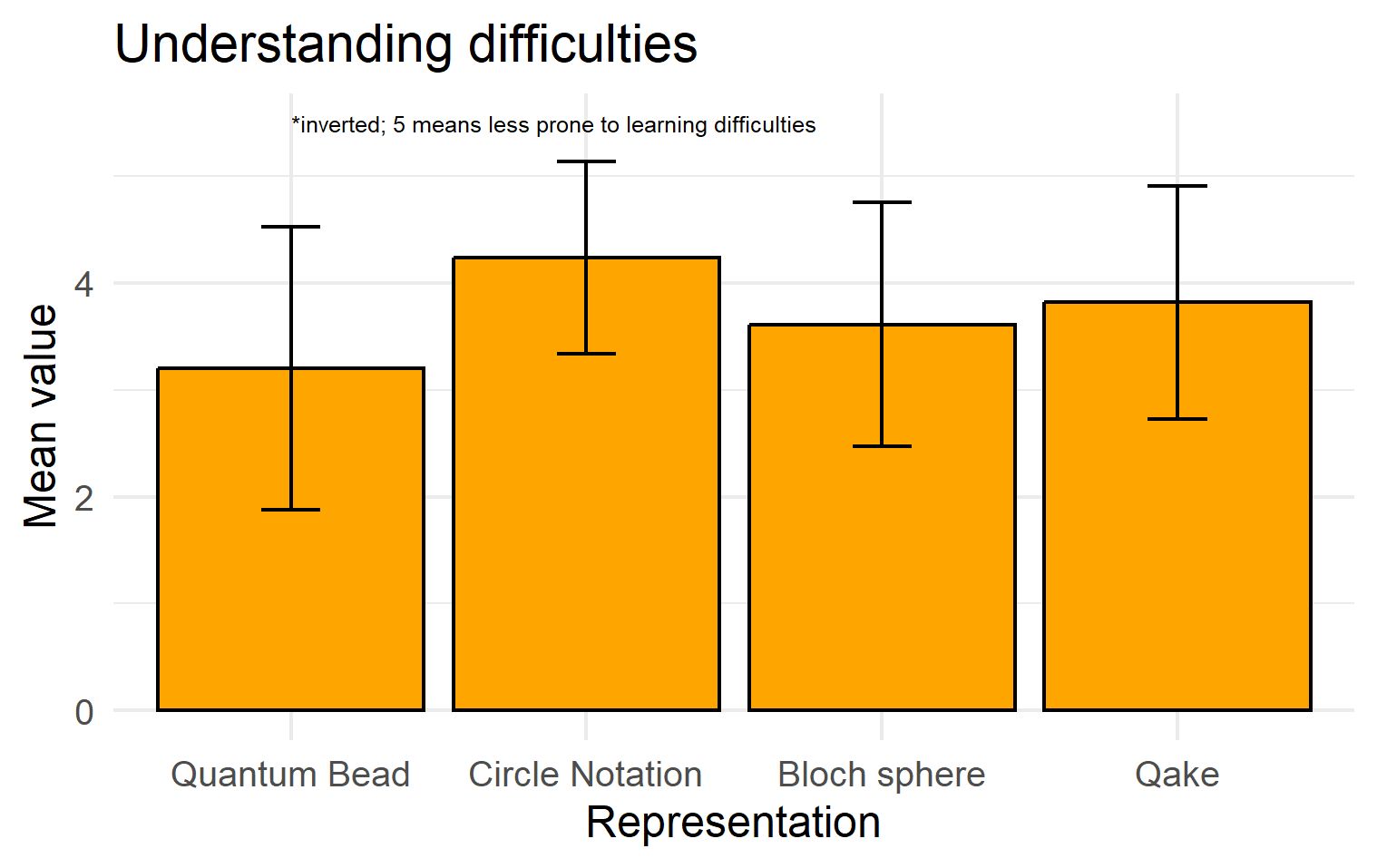}
        \caption{Understanding difficulties*}
        \tiny{*inverted: 5 means less prone to difficulties}
        \label{fig:mein_bild_3}
    \end{subfigure}
    \hfill
    \begin{subfigure}[b]{0.49\textwidth}
        \includegraphics[width=\textwidth]{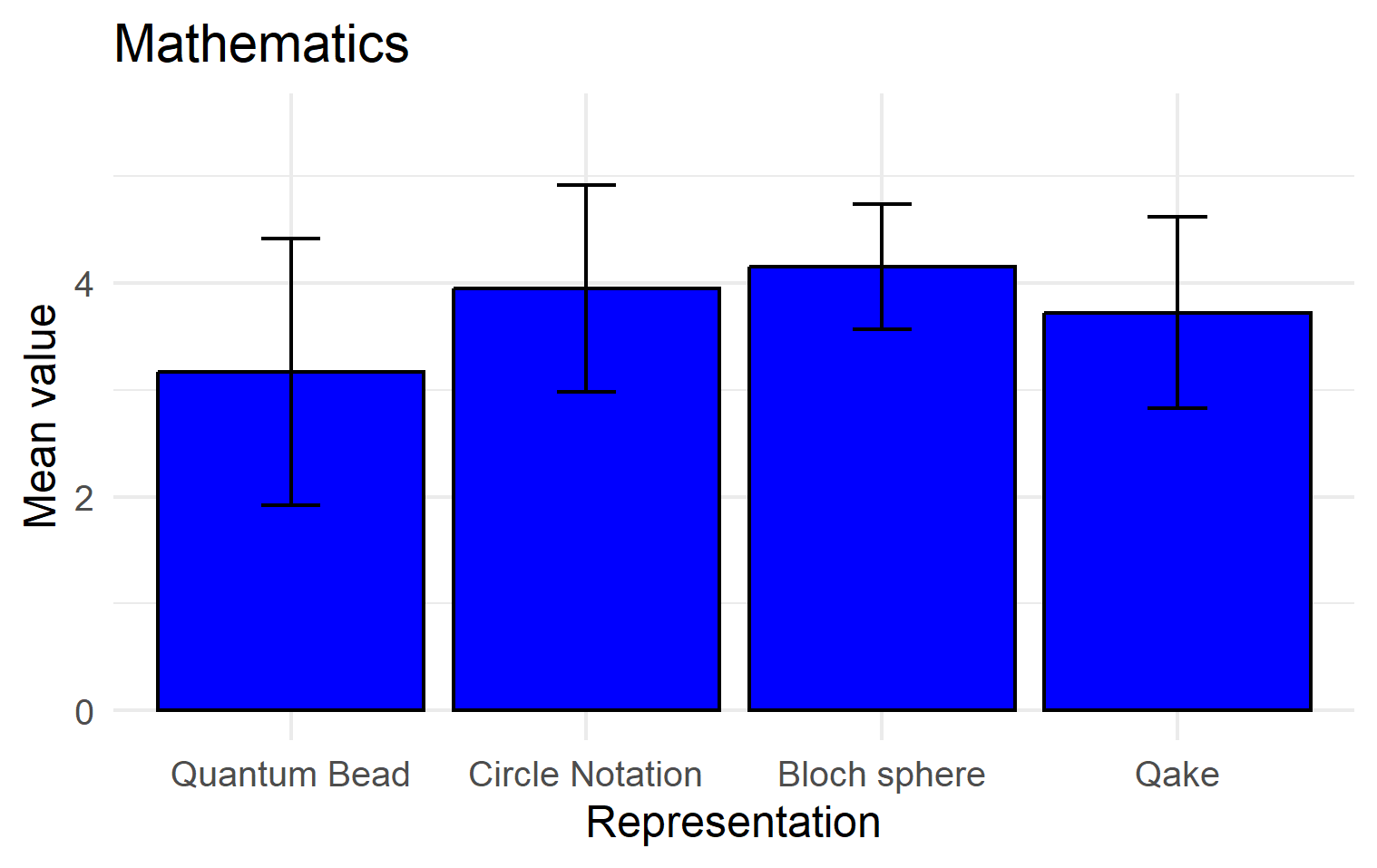}
        \caption{Mathematics}
        \label{fig:mein_bild_1.1}
    \end{subfigure}
    \hfill
    \begin{subfigure}[b]{0.49\textwidth}
        \includegraphics[width=\textwidth]{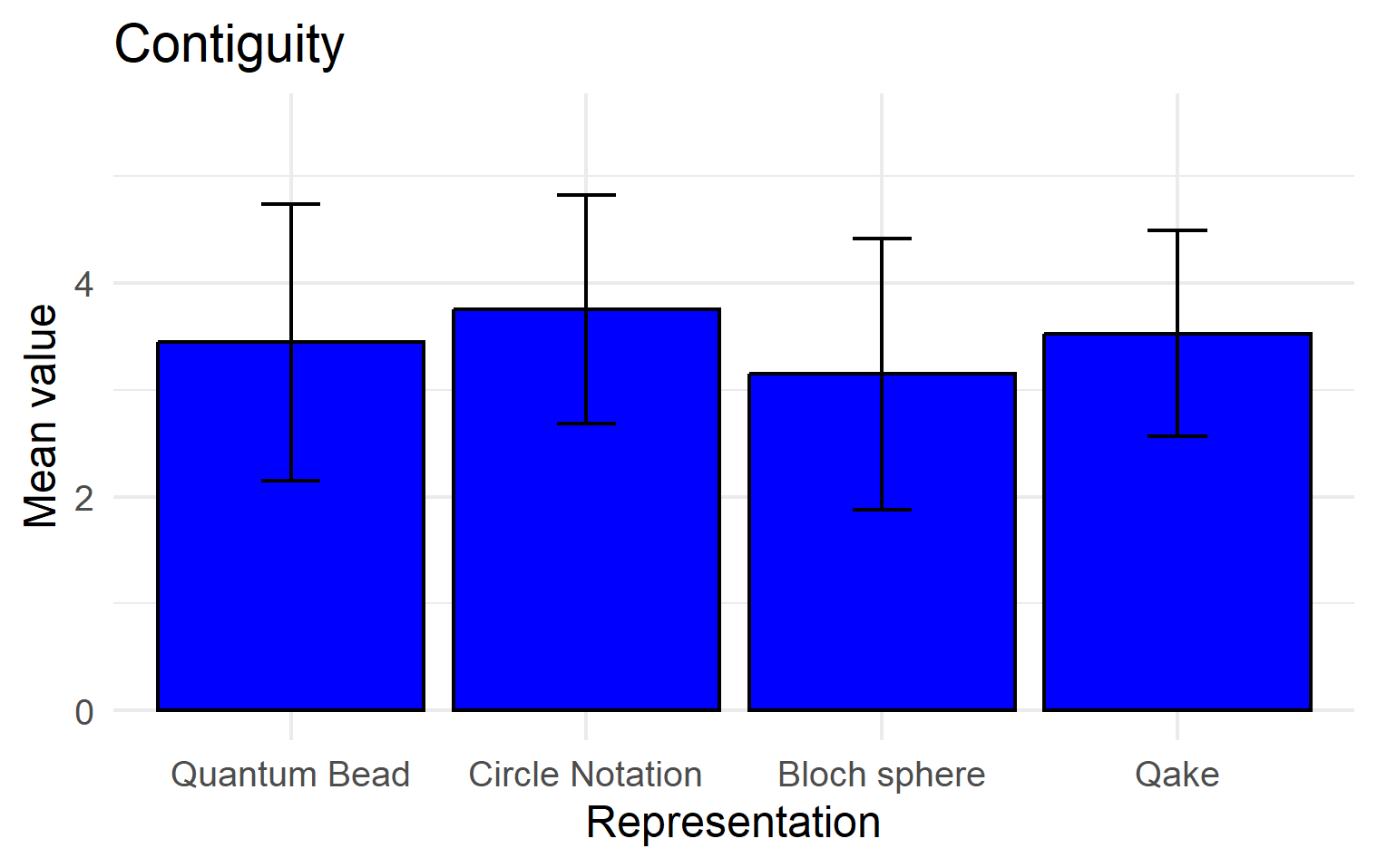}
        \caption{Contiguity}
        \label{fig:mein_bild_2.1}
    \end{subfigure}
    \hfill
    \begin{subfigure}[b]{0.49\textwidth}
        \includegraphics[width=\textwidth]{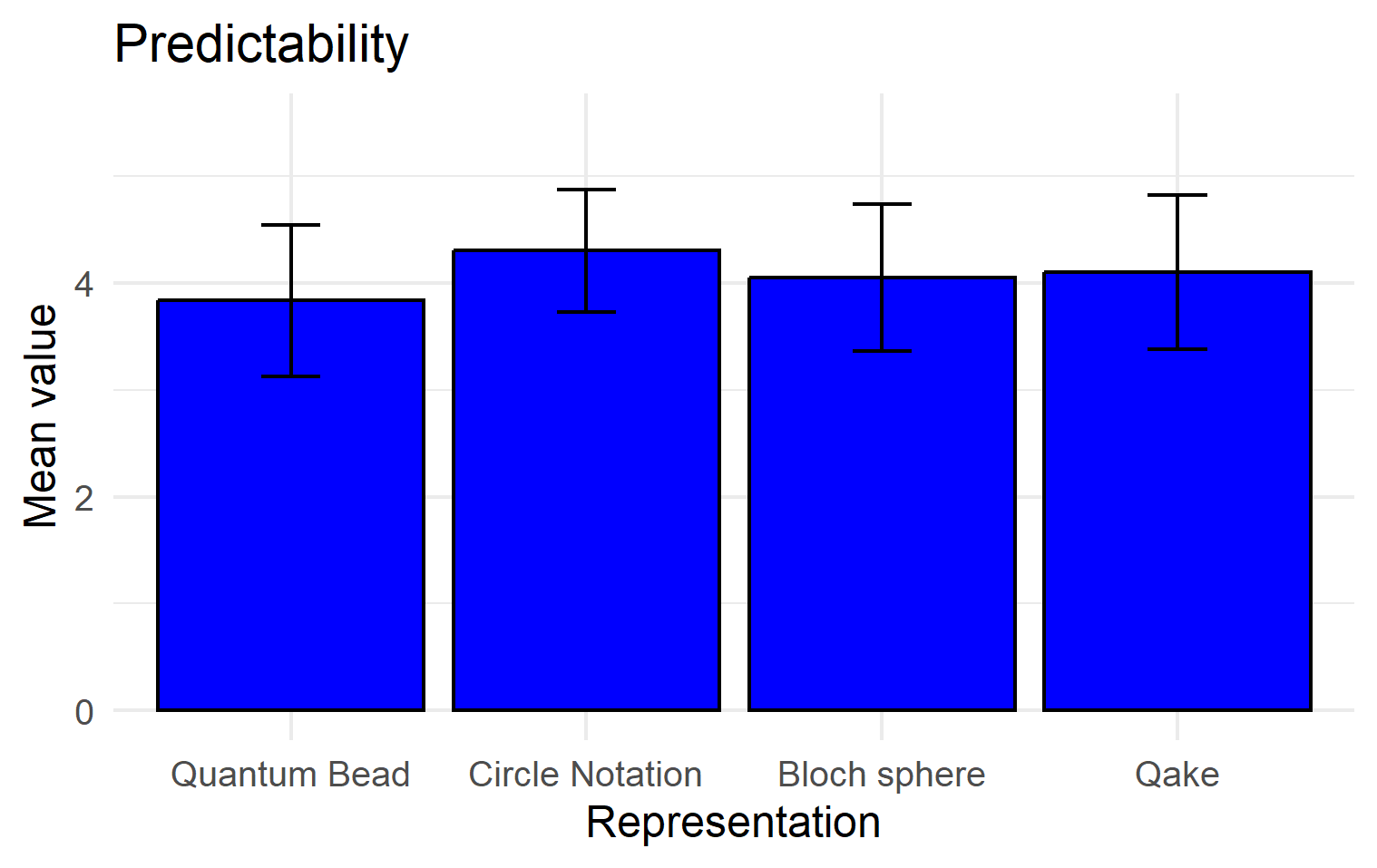}
        \caption{Predictability}
        \label{fig:mein_bild_3.1}
    \end{subfigure}
    \caption{Mean values for categories under design (a–c) and functions (d–f).}
    \label{fig:gesamtbild_1}
\end{figure}

\clearpage

\begin{figure}[H]
    \begin{subfigure}[b]{0.49\textwidth}
        \includegraphics[width=\textwidth]{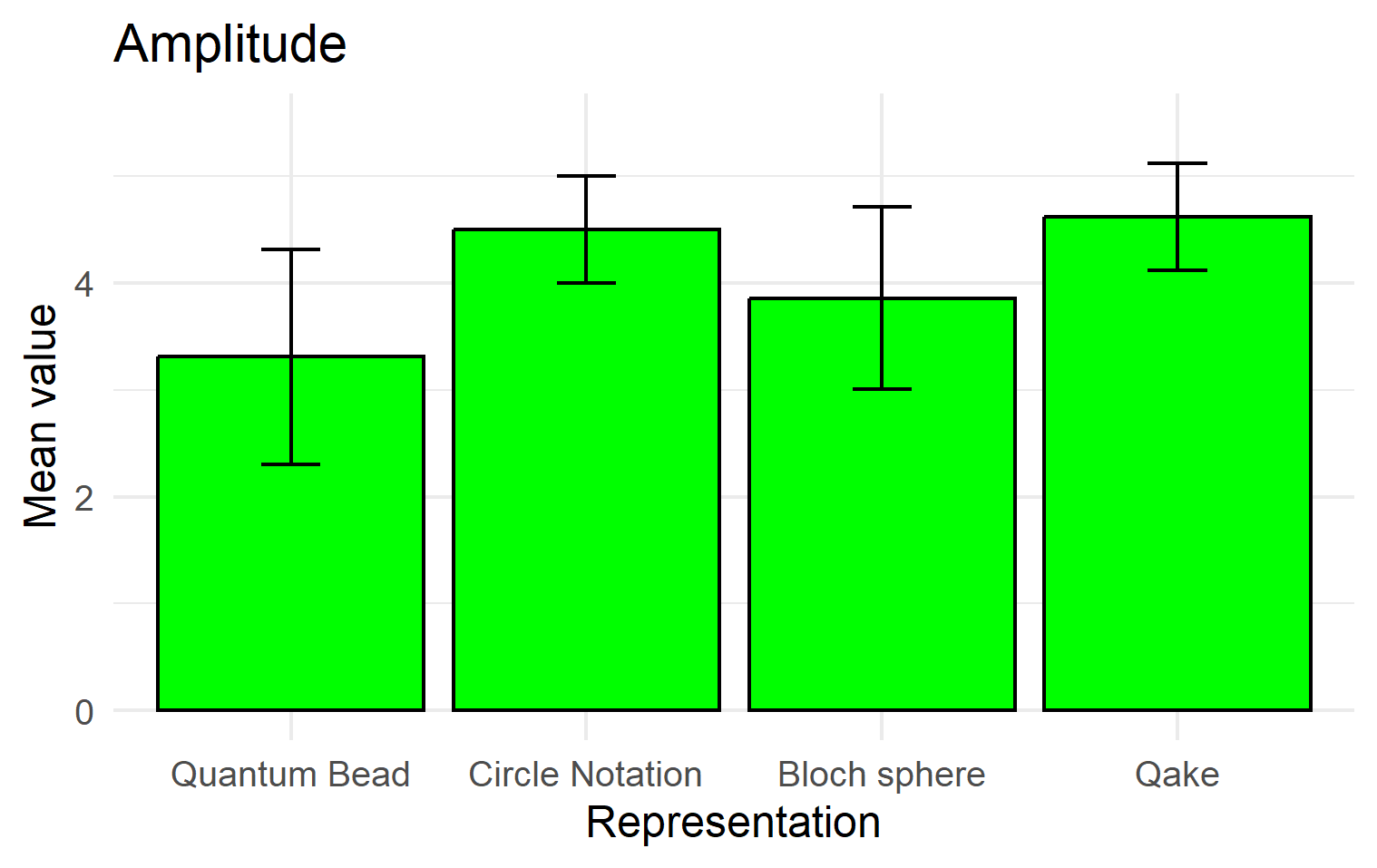}
        \caption{Amplitude}
        \label{fig:mein_bild_1.2}
    \end{subfigure}
    \hfill
    \begin{subfigure}[b]{0.49\textwidth}
        \includegraphics[width=\textwidth]{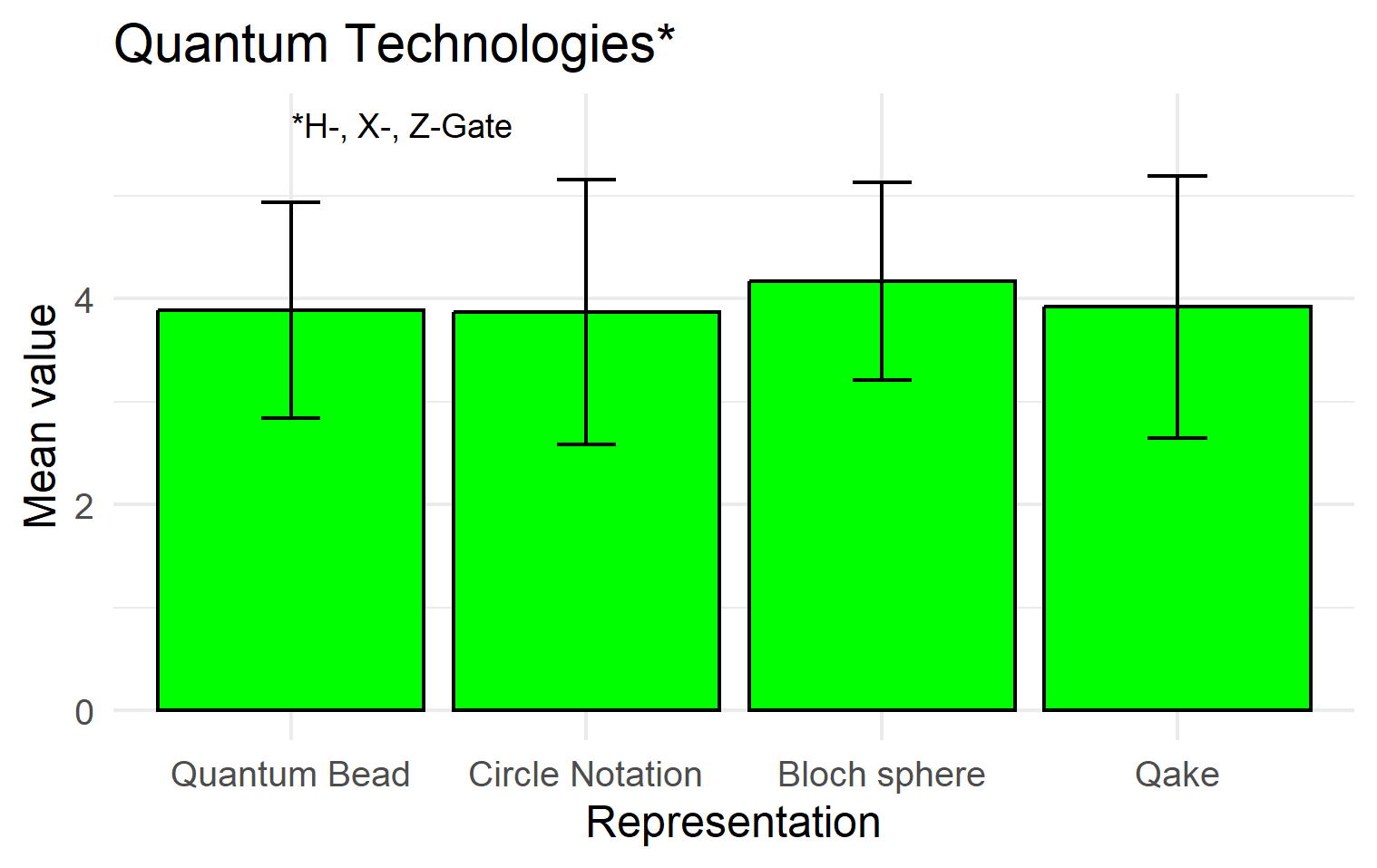}
        \caption{Quantum technologies}
        \label{fig:mein_bild_2.2}
    \end{subfigure}
    \hfill
    \begin{subfigure}[b]{0.49\textwidth}
        \includegraphics[width=\textwidth]{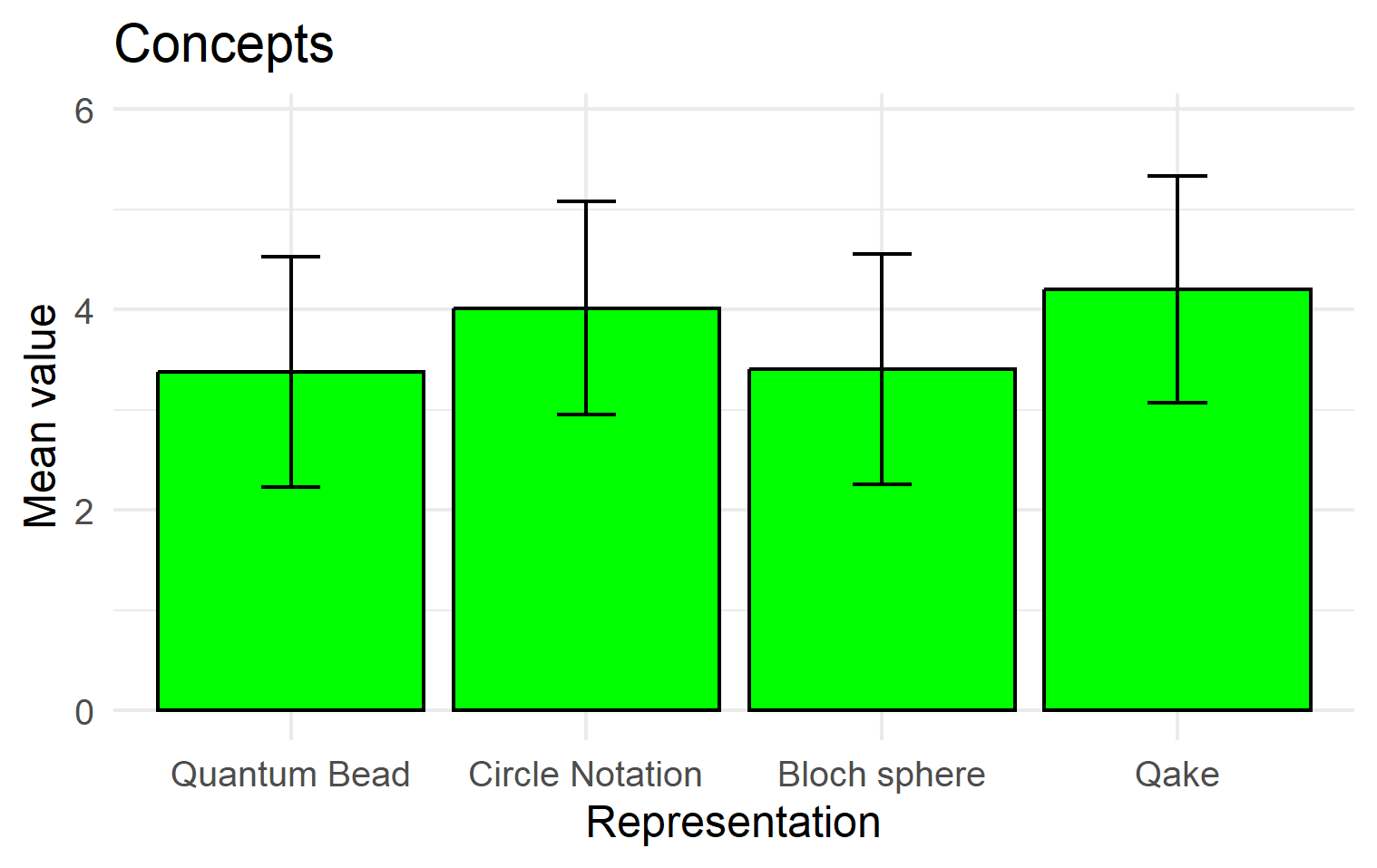}
        \caption{Concepts}
        \label{fig:mein_bild_3.2}
    \end{subfigure}
    \hfill
    \begin{subfigure}[b]{0.49\textwidth}
        \includegraphics[width=\textwidth]{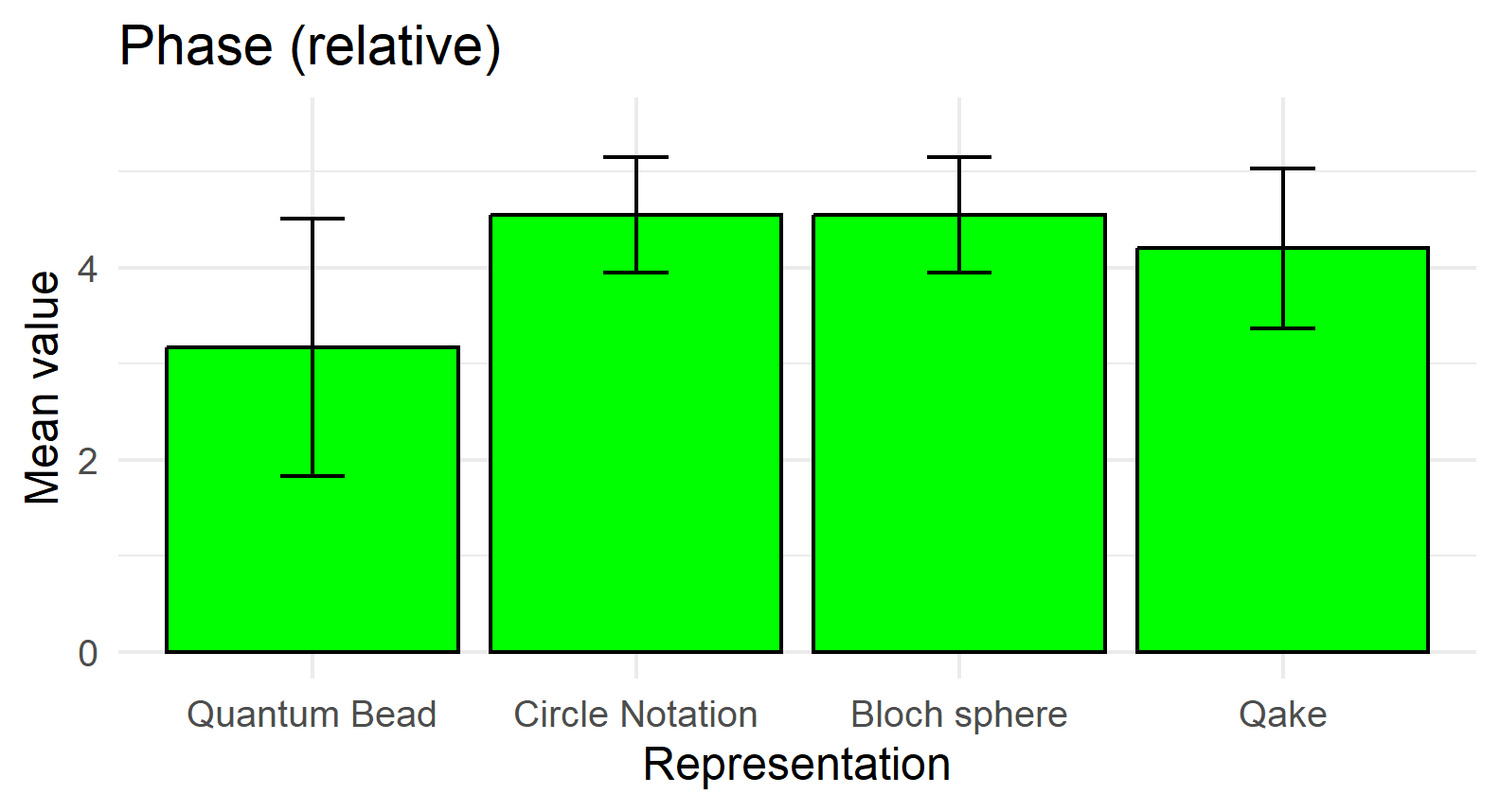}
        \caption{Phase}
        \label{fig:mein_bild_4.1}
    \end{subfigure}
    \caption{Mean values for categories under tasks (g–j).}
    \label{fig:gesamtbild_2}
\end{figure}
\end{appendices}


\newpage

\bibliography{references}

\begin{thebibliography}{10}
\providecommand{\doi}[1]{\url{https://doi.org/#1}}
\bibcommenthead

\bibitem[\protect\citeauthoryear{Passante}{2016}]{passante_energy_2016}
Passante G.
\newblock Energy measurement resources in spins-first and position-first quantum mechanics.
\newblock In: 2016 {Physics} {Education} {Research} {Conference} {Proceedings}. Sacramento, CA: American Association of Physics Teachers; 2016. p. 236--239.
\newblock Available from: \url{http://www.compadre.org/per/items/detail.cfm?ID=14237}.

\bibitem[\protect\citeauthoryear{Dür and Heusler}{2012}]{dur_was_2012}
Dür W, Heusler S.
\newblock Was man vom einzelnen {Qubit} über {Quantenphysik} lernen kann.
\newblock PhyDid A - Phys Didakt Sch Hochsch. 2012 Mar;1(11):01--16.
\newblock Number: 11.

\bibitem[\protect\citeauthoryear{Dür and Heusler}{2014}]{dur_was_2014}
Dür W, Heusler S.
\newblock Was man von zwei {Qubits} über {Quantenphysik} lernen kann: {Verschränkung} und {Quantenkorrelationen}.
\newblock PhyDid A - Phys Didakt Sch Hochsch. 2014 Jun;1(13):11--34.
\newblock Number: 13.

\bibitem[\protect\citeauthoryear{Greinert et~al.}{2023}]{greinert_future_2023}
Greinert F, Müller R, Bitzenbauer P, Ubben MS, Weber KA.
\newblock Future quantum workforce: competences, requirements, and forecasts.
\newblock Physical Review Physics Education Research. 2023 Jun;19(1):010137.
\newblock \doi{10.1103/PhysRevPhysEducRes.19.010137}.

\bibitem[\protect\citeauthoryear{Greinert and Müller}{2022}]{greinert_qualification_2022}
Greinert F, Müller R.
\newblock Qualification profiles for quantum technologies; 2022.
\newblock Available from: \url{https://zenodo.org/records/6834686}.

\bibitem[\protect\citeauthoryear{Sadaghiani}{2016}]{sadaghiani_spin_2016}
Sadaghiani HR.
\newblock Spin first vs. position first instructional approaches to teaching introductory quantum mechanics.
\newblock In: 2016 {Physics} {Education} {Research} {Conference} {Proceedings}. Sacramento, CA: American Association of Physics Teachers; 2016. p. 292--295.
\newblock Available from: \url{http://www.compadre.org/per/items/detail.cfm?ID=14251}.

\bibitem[\protect\citeauthoryear{Küblbeck and Müller}{2002}]{kublbeck2002wesenszuge}
Küblbeck J, Müller R.
\newblock Die {Wesenszüge} der {Quantenphysik}: {Modelle}, {Bilder}, {Experimente}.
\newblock Aulis-Verlag Deubner; 2002.

\bibitem[\protect\citeauthoryear{Müller and Mishina}{2021}]{muller_milqquantum_2021}
Müller R, Mishina O.
\newblock Milq—quantum physics in secondary school.
\newblock In: Jarosievitz B, Sükösd C, editors. Teaching-learning contemporary physics : from research to practice. Cham: Springer International Publishing; 2021. p. 33--45.
\newblock Available from: \url{https://doi.org/10.1007/978-3-030-78720-2_3}.

\bibitem[\protect\citeauthoryear{Müller and Greinert}{2024}]{MüllerGreinert+2024}
Müller R, Greinert F.
\newblock Quantum technologies.
\newblock Berlin, Boston: De Gruyter Oldenbourg; 2024.
\newblock Tex.title+duplicate-1: For Engineers.
\newblock Available from: \url{https://doi.org/10.1515/9783110717457}.

\bibitem[\protect\citeauthoryear{Merzel et~al.}{2024}]{merzel_core_2024}
Merzel A, Bitzenbauer P, Krijtenburg-Lewerissa K, Stadermann K, Andreotti E, Anttila D, et~al.
\newblock The core of secondary level quantum education: a multi-stakeholder perspective.
\newblock EPJ Quantum Technology. 2024 Dec;11(1):1--28.
\newblock \doi{10.1140/epjqt/s40507-024-00237-x}.

\bibitem[\protect\citeauthoryear{Sadaghiani and Munteanu}{2015}]{sadaghiani_spin_2015}
Sadaghiani HR, Munteanu J.
\newblock Spin first instructional approach to teaching quantum mechanics in sophomore level modern physics courses.
\newblock In: 2015 {Physics} {Education} {Research} {Conference} {Proceedings}. College Park, MD: American Association of Physics Teachers; 2015. p. 287--290.
\newblock Available from: \url{http://www.compadre.org/per/items/detail.cfm?ID=13956}.

\bibitem[\protect\citeauthoryear{Stadermann and van~den Berg}{2019}]{stadermann_analysis_2019}
Stadermann K, van~den Berg E.
\newblock Analysis of secondary school quantum physics curricula of 15 different countries: {Different} perspectives on a challenging topic.
\newblock Phys Rev Phys Educ Res. 2019 May;15(1):010130.
\newblock \doi{10.1103/PhysRevPhysEducRes.15.010130}.

\bibitem[\protect\citeauthoryear{Lemke}{}]{lemke_multiplying_nodate}
Lemke J.
\newblock Multiplying meaning.
\newblock Reading science: critical and functional perspectives on discourses of science;87.
\newblock London: Routledge London; 1998.

\bibitem[\protect\citeauthoryear{Manogue et~al.}{2012}]{manogue_representations_2012}
Manogue C, Gire E, McIntyre D, Tate J, Rebello NS, Engelhardt PV, et~al.
\newblock Representations for a spins-first approach to quantum mechanics.
\newblock Omaha, NE: AIP; 2012. p. 55--58.
\newblock Available from: \url{https://pubs.aip.org/aip/acp/article/1413/1/55-58/873092}.

\bibitem[\protect\citeauthoryear{Woitzik}{2020}]{woitzik_quanteninformationsverarbeitung_2020}
Woitzik AJ.
\newblock Quanteninformationsverarbeitung in der gymnasialen {Oberstufe}.
\newblock arXiv:201101029 [preprind][cited 2024 Sep 13]. 2020;[cited 2024 Sep 13].

\bibitem[\protect\citeauthoryear{Feynman}{2006}]{feynman2006qed}
Feynman RP.
\newblock {QED}: the strange theory of light and matter.
\newblock Princeton, NJ: Princeton University Press; 2006.

\bibitem[\protect\citeauthoryear{Bader}{1994}]{bader1994optik}
Bader F.
\newblock Optik und {Quantenphysik} nach {Feynmans} {QED}.
\newblock Phys Sch. 1994;32(7):8.

\bibitem[\protect\citeauthoryear{Küblbeck}{1997}]{kublbeck1997modellbildung}
Küblbeck J.
\newblock Modellbildung in der {Physik}.
\newblock Suttgart: Landesinstitut für Schulentwicklung,. 1997;.

\bibitem[\protect\citeauthoryear{Küblbeck}{2015}]{kublbeck_quantenphysik_2015}
Küblbeck J.
\newblock Quantenphysik.
\newblock In: Kircher E, Girwidz R, Häußler P, editors. Physikdidaktik: {Theorie} und {Praxis}. Berlin, Heidelberg: Springer; 2015. p. 479--501.
\newblock Available from: \url{doi: 10.1007/978-3-642-41745-0_15}.

\bibitem[\protect\citeauthoryear{Johnston et~al.}{2019}]{johnston_programming_2019}
Johnston ER, Harrigan N, Gimeno-Segovia M.
\newblock Programming quantum computers: essential algorithms and code samples.
\newblock Sebastopol, CA: O'Reilly Media, Inc.; 2019.

\bibitem[\protect\citeauthoryear{Bley et~al.}{2024}]{bley_visualizing_2024}
Bley J, Rexigel E, Arias A, Longen N, Krupp L, Kiefer-Emmanouilidis M, et~al.
\newblock Visualizing entanglement in multiqubit systems.
\newblock Phys Rev Res. 2024 Apr;6(2):023077.
\newblock \doi{10.1103/PhysRevResearch.6.023077}.

\bibitem[\protect\citeauthoryear{Just}{2020}]{just_quantencomputing_2020}
Just B.
\newblock Quantencomputing kompakt: {Spukhafte} {Fernwirkung} und {Teleportation} endlich verständlich.
\newblock {IT} kompakt. Berlin, Heidelberg: Springer; 2020.
\newblock Available from: \url{https://link.springer.com/10.1007/978-3-662-61889-9}.

\bibitem[\protect\citeauthoryear{Garon et~al.}{2015}]{garon_visualizing_2015}
Garon A, Zeier R, Glaser SJ.
\newblock Visualizing operators of coupled spin systems.
\newblock Phys Rev A. 2015 Apr;91(4):042122.
\newblock \doi{10.1103/PhysRevA.91.042122}.

\bibitem[\protect\citeauthoryear{Yeung}{2020}]{yeung2020quantum}
Yeung K.
\newblock Quantum computing \& some physics: the quantum computing comics notebook.
\newblock Verlag nicht ermittelbar; 2020.
\newblock Available from: \url{https://books.google.de/books?id=3LEVzgEACAAJ}.

\bibitem[\protect\citeauthoryear{Donhauser et~al.}{2024}]{Donhauser2024}
Donhauser A, Kuhn J, {et al }.
\newblock Qubit cake model; 2024.

\bibitem[\protect\citeauthoryear{Weidner et~al.}{2021}]{weidner2021publications}
Weidner CA, Ahmed SZ, Jensen JH, Sherson JF.: Publications using {Quatomic} software.

\bibitem[\protect\citeauthoryear{Küchemann et~al.}{2023}]{kuchemann_impact_2023}
Küchemann S, Ubben M, Dzsotjan D, Mukhametov S, Weidner CA, Qerimi L, et~al.: The impact of an interactive visualization and simulation tool on learning quantum physics: results of an eye-tracking study.
\newblock ArXiv:2302.06286 [preprint]. [cited 2024 Sep 13].

\bibitem[\protect\citeauthoryear{Kohnle et~al.}{2020}]{kohnle_sketching_2020}
Kohnle A, Ainsworth SE, Passante G.
\newblock Sketching to support visual learning with interactive tutorials.
\newblock Phys Rev Phys Educ Res. 2020 Dec;16(2):020139.
\newblock \doi{10.1103/PhysRevPhysEducRes.16.020139}.

\bibitem[\protect\citeauthoryear{Kohnle et~al.}{2015}]{kohnle_investigating_2015}
Kohnle A, Baily C, Ruby S.
\newblock Investigating the influence of visualization on student understanding of quantum superposition.
\newblock In: 2014 {Physics} {Education} {Research} {Conference} {Proceedings}. Minneapolis, MS: American Association of Physics Teachers; 2015. p. 139--142.
\newblock Available from: \url{http://www.compadre.org/per/items/detail.cfm?ID=13468}.

\bibitem[\protect\citeauthoryear{Goorney et~al.}{2023}]{goorney_quantum_2023}
Goorney S, Bley J, Heusler S, Sherson J.: The {Quantum} {Curriculum} {Transformation} {Framework} for the development of {Quantum} {Information} {Science} and {Technology} {Education}.
\newblock arXiv.
\newblock ArXiv:2308.10371 [physics, physics:quant-ph].
\newblock Available from: \url{http://arxiv.org/abs/2308.10371}.

\bibitem[\protect\citeauthoryear{Rau}{2017}]{rau_conditions_2017}
Rau MA.
\newblock Conditions for the effectiveness of multiple visual representations in enhancing {STEM} learning.
\newblock Educ Psychol Rev. 2017 Dec;29(4):717--761.
\newblock \doi{10.1007/s10648-016-9365-3}.

\bibitem[\protect\citeauthoryear{Garcia~Garcia and Cox}{2010}]{garcia_garcia_graph-as-picture_2010}
Garcia~Garcia G, Cox R.
\newblock “{Graph}-as-{Picture}” misconceptions in young students.
\newblock In: Goel AK, Jamnik M, Narayanan NH, editors. Diagrammatic representation and inference. Berlin, Heidelberg: Springer; 2010. p. 310--312.

\bibitem[\protect\citeauthoryear{Wiesner}{1996}]{wiesner1996verstandnisse}
Wiesner H.
\newblock Verständnisse von {Leistungskursschülern} über {Quantenphysik}.
\newblock Phys Sch. 1996;34:95.

\bibitem[\protect\citeauthoryear{Fischler and Lichtfeldt}{1992}]{fischler_modern_1992}
Fischler H, Lichtfeldt M.
\newblock Modern physics and students’ conceptions.
\newblock Int J Sci Educ. 1992 Apr;14(2):181--190.
\newblock \doi{10.1080/0950069920140206}.

\bibitem[\protect\citeauthoryear{Singh and Marshman}{2015}]{singh_review_2015}
Singh C, Marshman E.
\newblock Review of student difficulties in upper-level quantum mechanics.
\newblock Phys Rev Special Top Phys Educ Res. 2015 Sep;11(2):020117.
\newblock \doi{10.1103/PhysRevSTPER.11.020117}.

\bibitem[\protect\citeauthoryear{Özcan}{2011}]{ozcan_what_2011}
Özcan O.
\newblock What are the students’ mental models about the "spin" and "photon" concepts in modern physics?
\newblock Procedia - Soc Behav Sci. 2011 Jan;15:1372--1375.
\newblock \doi{10.1016/j.sbspro.2011.03.295}.

\bibitem[\protect\citeauthoryear{Henriksen et~al.}{2018}]{henriksen_what_2018}
Henriksen EK, Angell C, Vistnes AI, Bungum B.
\newblock What is light? {Students}’ reflections on the wave-particle duality of light and the nature of physics.
\newblock Sci \& Educ. 2018;27:81--111.

\bibitem[\protect\citeauthoryear{Ubben and Bitzenbauer}{2022}]{ubben_two_2022}
Ubben MS, Bitzenbauer P.
\newblock Two cognitive dimensions of students’ mental models in science: fidelity of gestalt and functional fidelity.
\newblock Educ Sci. 2022 Mar;12(3):163.
\newblock \doi{10.3390/educsci12030163}.

\bibitem[\protect\citeauthoryear{Schnotz}{2001}]{schnotz_kognitive_2001}
Schnotz W.
\newblock Kognitive {Prozesse} bei der sprach- und bildgestützten {Konstruktion} mentaler {Modelle}.
\newblock In: Sichelschmidt L, Strohner H, editors. Sprache, {Sinn} und {Situation}: {Festschrift} für {Gert} {Rickheit} zum 60. {Geburtstag}. Wiesbaden: Deutscher Universitätsverlag; 2001. p. 43--57.

\bibitem[\protect\citeauthoryear{Kosslyn}{1989}]{kosslyn_understanding_1989}
Kosslyn SM.
\newblock Understanding charts and graphs.
\newblock Appl Cogn Psychol. 1989;3(3):185--225.
\newblock \doi{10.1002/acp.2350030302}.

\bibitem[\protect\citeauthoryear{Bertin}{1983}]{bertin1983semiology}
Bertin J.
\newblock Semiology of graphics.
\newblock University of Wisconsin Press; 1983.

\bibitem[\protect\citeauthoryear{Ainsworth}{2006}]{ainsworth_deft_2006}
Ainsworth S.
\newblock {DeFT}: {A} conceptual framework for considering learning with multiple representations.
\newblock Learn Instr. 2006 Jun;16(3):183--198.
\newblock \doi{10.1016/j.learninstruc.2006.03.001}.

\bibitem[\protect\citeauthoryear{Huber et~al.}{2024}]{Huber2024}
Huber D, Glaser S, {et al }.
\newblock Quantum bead; 2024.
\newblock Munich: Technical University of Munich.

\bibitem[\protect\citeauthoryear{Leiner et~al.}{2017}]{leiner_wigner_2017}
Leiner D, Zeier R, Glaser SJ.
\newblock Wigner tomography of multispin quantum states.
\newblock Phys Rev A. 2017 Dec;96(6):063413.
\newblock \doi{10.1103/PhysRevA.96.063413}.

\bibitem[\protect\citeauthoryear{Dür and Heusler}{2014}]{dur_visualization_2014}
Dür W, Heusler S.
\newblock Visualization of the invisible: the qubit as key to quantum physics.
\newblock Phys Teach. 2014 Nov;52(8):489--492.
\newblock \doi{10.1119/1.4897588}.

\bibitem[\protect\citeauthoryear{Hewstone and Stroebe}{2021}]{hewstone2021introduction}
Hewstone M, Stroebe W.
\newblock An introduction to social psychology.
\newblock Hoboken, NJ: John Wiley \& Sons; 2021.

\bibitem[\protect\citeauthoryear{van Gog}{2014}]{mayer_signaling_2014}
van Gog T.
\newblock The {Signaling} (or cueing) {Principle} in {Multimedia} {Learning}.
\newblock In: Mayer RE, editor. In: {The} {Cambridge} {Handbook} of {Multimedia} {Learning}. 2nd ed. Cambridge {Handbooks} in {Psychology}. Cambridge: Cambridge University Press; 2014. p. 263--278.
\newblock Available from: \url{DOI: 10.1017/CBO9781139547369.014}.

\bibitem[\protect\citeauthoryear{Castro-Alonso and Jansen}{2019}]{castro-alonso_sex_2019}
Castro-Alonso JC, Jansen P.
\newblock Sex differences in visuospatial processing.
\newblock In: Castro-Alonso JC, editor. In: {Visuospatial} processing for education in health and natural sciences. Cham: Springer International Publishing; 2019. p. 81--110.

\bibitem[\protect\citeauthoryear{Heo and Toomey}{2020}]{heo_learning_2020}
Heo M, Toomey N.
\newblock Learning with multimedia: the effects of gender, type of multimedia learning resources, and spatial ability.
\newblock Computers \& Education. 2020 Mar;146:103747.
\newblock \doi{10.1016/j.compedu.2019.103747}.

\bibitem[\protect\citeauthoryear{Saha and Halder}{2016}]{saha_he_2016}
Saha S, Halder S.: He or she: does gender affect various modes of instructional visual design?
\newblock Texas State University, Center for Diversity and Gender Studies.

\bibitem[\protect\citeauthoryear{Sweller}{1994}]{sweller_cognitive_1994}
Sweller J.
\newblock Cognitive load theory, learning difficulty, and instructional design.
\newblock Learn Instr. 1994 Jan;4(4):295--312.
\newblock \doi{10.1016/0959-4752(94)90003-5}.

\bibitem[\protect\citeauthoryear{Baddeley}{1992}]{baddeley_working_1992}
Baddeley A.
\newblock Working memory.
\newblock Science. 1992;255(5044):556--559.

\bibitem[\protect\citeauthoryear{Miller}{1956}]{miller_magical_1956}
Miller GA.
\newblock The magical number seven, plus or minus two: some limits on our capacity for processing information.
\newblock Psychol Rev. 1956;63(2):81.

\bibitem[\protect\citeauthoryear{Sweller et~al.}{2019}]{sweller_cognitive_2019}
Sweller J, van Merriënboer JJG, Paas F.
\newblock Cognitive architecture and instructional design: 20 years later.
\newblock Educ Psychol Rev. 2019 Jun;31(2):261--292.
\newblock \doi{10.1007/s10648-019-09465-5}.

\bibitem[\protect\citeauthoryear{Schrödinger}{1935}]{schrodinger_gegenwartige_1935}
Schrödinger E.
\newblock Die gegenwärtige {Situation} in der {Quantenmechanik}.
\newblock Naturwissenschaften. 1935;23(50):844--849.

\bibitem[\protect\citeauthoryear{Mayer and Fiorella}{2014}]{Mayer_Fiorella_2014}
Mayer RE, Fiorella L.
\newblock Principles for reducing extraneous processing in multimedia learning: coherence, signaling, redundancy, spatial contiguity, and temporal contiguity principles.
\newblock In: Mayer REE, editor. The {Cambridge} handbook of multimedia learning. Cambridge: Cambridge University Press; 2014. p. 279--315.

\bibitem[\protect\citeauthoryear{Ott et~al.}{2018}]{ott_multiple_2018}
Ott N, Brünken R, Vogel M, Malone S.
\newblock Multiple symbolic representations: the combination of formula and text supports problem solving in the mathematical field of propositional logic.
\newblock Learn Instr. 2018 Dec;58:88--105.
\newblock \doi{10.1016/j.learninstruc.2018.04.010}.

\bibitem[\protect\citeauthoryear{Krüger et~al.}{2018}]{kruger_modelle_2018}
Krüger D, Kauertz A, Upmeier~zu Belzen A.
\newblock Modelle und das {Modellieren} in den {Naturwissenschaften}.
\newblock In: Krüger D, Parchmann I, Schecker H, editors. Theorien in der naturwissenschaftsdidaktischen {Forschung}. Berlin, Heidelberg: Springer; 2018. p. 141--157.
\newblock Available from: \url{https://doi.org/10.1007/978-3-662-56320-5_9}.

\bibitem[\protect\citeauthoryear{Treagust et~al.}{2017}]{treagust_multiple_2017}
Treagust DF, Duit R, Fischer HE, editors.
\newblock Multiple representations in physics education. vol.~10.
\newblock Cham: Springer International Publishing; 2017.
\newblock Available from: \url{http://link.springer.com/10.1007/978-3-319-58914-5}.

\bibitem[\protect\citeauthoryear{Nielsen and Chuang}{2010}]{nielsen_quantum_2010}
Nielsen MA, Chuang IL.
\newblock Quantum computation and quantum information: 10th anniversary edition.
\newblock Cambridge: Cambridge University Press; 2010.

\bibitem[\protect\citeauthoryear{Feynman et~al.}{2010}]{feynman_quantum_2010}
Feynman RP, Hibbs AR, Styer DF.
\newblock Quantum mechanics and path integrals.
\newblock Courier Corporation; 2010.

\bibitem[\protect\citeauthoryear{Sweller}{1988}]{sweller_cognitive_1988}
Sweller J.
\newblock Cognitive load during problem solving: effects on learning.
\newblock Cogn Sci. 1988;12(2):257--285.

\bibitem[\protect\citeauthoryear{Marshman and Singh}{2017}]{marshman_investigating_2017}
Marshman E, Singh C.
\newblock Investigating and improving student understanding of quantum mechanics in the context of single photon interference.
\newblock Phys Rev Phys Educ Res. 2017 Apr;13(1):010117.
\newblock \doi{10.1103/PhysRevPhysEducRes.13.010117}.

\bibitem[\protect\citeauthoryear{Ubben and Heusler}{2021}]{ubben_gestalt_2021}
Ubben MS, Heusler S.
\newblock Gestalt and functionality as independent dimensions of mental models in science.
\newblock Res Sci Educ. 2021 Oct;51(5):1349--1363.
\newblock \doi{10.1007/s11165-019-09892-y}.

\bibitem[\protect\citeauthoryear{Scholz et~al.}{2006}]{scholz_deutsch-jozsa_2006}
Scholz M, Aichele T, Ramelow S, Benson O.
\newblock Deutsch-{Jozsa} algorithm using triggered single photons from a single quantum dot.
\newblock Physical Review Letters. 2006 May;96(18):180501.
\newblock \doi{10.1103/PhysRevLett.96.180501}.

\bibitem[\protect\citeauthoryear{O'Brien et~al.}{2003}]{obrien_demonstration_2003}
O'Brien JL, Pryde GJ, White AG, Ralph TC, Branning D.
\newblock Demonstration of an all-optical quantum controlled-{NOT} gate.
\newblock Nature. 2003 Nov;426(6964):264--267.
\newblock \doi{10.1038/nature02054}.

\bibitem[\protect\citeauthoryear{Patton}{2014}]{patton_qualitative_2014}
Patton MQ.
\newblock Qualitative research \& evaluation methods: integrating theory and practice.
\newblock SAGE Publications; 2014.
\newblock Google-Books-ID: ovAkBQAAQBAJ.

\bibitem[\protect\citeauthoryear{Zinn et~al.}{2001}]{zinn_identifying_2001}
Zinn J, Zalokowski A, Hunter L.
\newblock Identifying indicators of laboratory management performance: a multiple constituency approach.
\newblock Health Care Manag Rev. 2001;26(1):40.

\bibitem[\protect\citeauthoryear{von~der Gracht}{2012}]{von_der_gracht_consensus_2012}
von~der Gracht HA.
\newblock Consensus measurement in {Delphi} studies: review and implications for future quality assurance.
\newblock Technol Forecast and Soc Change. 2012 Oct;79(8):1525--1536.
\newblock \doi{10.1016/j.techfore.2012.04.013}.

\bibitem[\protect\citeauthoryear{Field et~al.}{2012}]{field_discovering_2012}
Field A, Field Z, Miles J.
\newblock Discovering statistics using {R}; 2012.

\bibitem[\protect\citeauthoryear{Cohen}{1988}]{cohen_statistical_1988}
Cohen J.
\newblock Statistical power analysis for the behavioral sciences.
\newblock 2nd ed. Hillsdale, N.J: L. Erlbaum Associates; 1988.

\end{thebibliography}

\end{document}